# Direct electrical access to the spin manifolds of individual monovalent lanthanide atoms

G. Czap[1], K. Noh[1,2,3], J. Velasco Jr.[1,4], R. M. Macfarlane[1], H. Brune[1,5]* and C. P. Lutz[1]*

[1]IBM Almaden Research Center, 650 Harry Road, San Jose, CA 95120, USA

[2]Center for Quantum Nanoscience (QNS), Institute of Basic Science (IBS), 03760 Seoul, Republic of Korea

[3]Department of Physics, Ewha Womans University, 03760 Seoul, Republic of Korea

[4]Department of Physics, University of California, Santa Cruz CA 95064, USA

[5]Institute of Physics, École Polytechnique Fédérale de Lausanne (EPFL), CH-1015 Lausanne, Switzerland

*Corresponding authors. Email: cplutz@us.ibm.com, harald.brune@epfl.ch

**Abstract:** Lanthanide atoms show long magnetic lifetimes because of their strongly localized 4*f* electrons, but electrical control of their spins has been difficult because of their closed valence shell configurations. We achieved electron spin resonance of individual lanthanide atoms using a scanning tunneling microscope to probe the atoms bound to a protective insulating film. These atoms were prepared in the monovalent state with an unpaired 6*s* electron, enabling tunnel current to access their 4*f* electrons. Europium spectra display a rich array of transitions among the 54 combined electron and nuclear spin states. In contrast, samarium's ground state is a Kramers doublet with an extraordinarily large *g*-factor of nearly 5. These results demonstrate that all-electronic sensing and control of individual lanthanide spins is possible for quantum devices and spin-based electronics by using their rarely-observed monovalent cation state.

Lanthanide atoms combine localized 4*f* electron spins with strong spin-orbit coupling, making them highly promising for single-ion magnetic memories and quantum bits. They exhibit magnetic bistability in single-ion molecular and atomic magnets[1–3], and very long coherence times in molecules[4–6] or when dilutely dissolved in ionic crystals[7–10]. Single lanthanide ions also offer efficient light–matter–spin interactions for use in quantum communication[9,11]. Electron spin resonance (ESR) is a versatile technique capable of coherently controlling and addressing the properties of quantum spin systems, including access to hyperfine interactions between electron and nuclear spins[12]. Combining ESR with scanning tunneling microscopy (STM) allows this information to be obtained for individual atoms, molecules, and assembled nanostructures in an environment controlled on the atomic scale [13]. Thus far, spins in 3*d* atoms[13–20] and in spin-$\frac{1}{2}$ molecules[21–23] have been addressed in ESR-STM. In contrast, no direct resonant driving and sensing of lanthanide spins has been reported, although their spins can be accessed indirectly through the use of a neighboring 3*d* adatom[24].

Here we demonstrate that the open valence shell character of monovalent lanthanide atoms allows direct electrical driving and sensing of the 4*f* atomic spins. Monovalent lanthanides Eu(I) and Sm(I) have been proposed to exist as charged dopants in irradiated[25,26] or additively reduced[26] alkali metal halide crystals. Other monovalent lanthanides were observed only recently, in borozene complexes[27] and in lanthanides on metal-supported graphene[28]. The monovalent electron configuration is otherwise very uncommon for lanthanides[27,29]. We obtained the monovalent state by choosing elements that occur readily as divalent species in compounds where they retain their full complement of 4*f* electrons, which include Eu and Sm[30]. We stabilized the Sm and Eu cations in the monovalent state by adsorbing them on a thin film of insulating MgO grown on an Ag(100) surface, where they spontaneously ionize to the +1 charge state having the open $6s^1$ valence configuration.

We present exceptionally rich ESR spectra of individual $Eu^+$ and $Sm^+$ ions. These ESR spectra are complemented by inelastic electron tunneling spectra (IETS) to probe the higher-energy magnetic excitations that allow determination of the electronic configurations. The *f*-shell occupations are half-full ($f^7$ for Eu) or nearly half-full ($f^6$ for Sm) which results in high *f*-spin configurations for both elements, but with markedly different properties. Eu has a large total spin $S = 4$ and vanishing orbital angular momentum. This enables access to transitions between crystal-field split states for the first time with ESR-STM. The ESR spectra of Eu show at least four groups of six hyperfine-split peaks because of its large manifold of electronic and nuclear spin states and its unusually small magneto-crystalline anisotropy. In contrast, Sm features a Kramers doublet ground state with nearly canceling spin and orbital angular momentum, resulting in an unusually large and anisotropic g-factor close to 5. For each odd isotope of both elements, the ESR spectra reveal the hyperfine interaction $A$, electron g-factor, and for Eu, the magneto-crystalline anisotropy parameters. These results show that individual lanthanide atomic spins with an open-shell valence configuration can be directly sensed and controlled electrically.

Atoms of both elements were found to bind at three different sites on MgO: at the oxygen site (where they appear ~0.3 nm tall), oxygen-oxygen bridge site (~0.3 nm) and magnesium site (~0.2 nm) (Fig. 1b,c and Fig. S1). In contrast to other lanthanide elements on MgO thin films[31–33], we find that Eu and Sm have a preference to adsorb on bridge and Mg sites. We were able to readily reposition them between the bridge or Mg sites, and for some tips, the oxygen site, using atomic manipulation (see Methods).

## Europium spin excitations

When Eu was adsorbed on the bridge site (designated Eu-b) it exhibited IETS excitations seen as steps in the differential conductance (d$I$/d$V$) at 163 mV (Fig. 1e). This excitation shifted with magnetic field (Fig. 1e) and showed sensitivity to the spin polarization of the tip[34] (Fig. S2e), indicating that it is a magnetic excitation of the atom. Comparison with optical spectroscopy data for gas phase Eu ions[35], and further measurements below, indicate that the configuration of Eu-b is the one-fold positively charged ion Eu$^+$ with a ground-state configuration of $4f^7\,5d^0\,6s^1$, which has an open valence shell as desired. Its electronic state has f-shell spin $S_{4f} = 7/2$, orbital angular momentum $L_{4f} = 0$, total f-shell angular momentum $J_{4f} = 7/2$, valence (s-shell) electron spin $S_{val} = 1/2$, total spin $S_{tot} = 4$, and total angular momentum $J_{tot} = S_{tot} = 4$. The 4f spin is aligned with the valence spin, giving the electronic term $^9S_4$. The lowest excitation of gas-phase Eu$^+$ makes a transition to $^7S_3$, which corresponds to an intra-atomic spin flip of the 6s valence electron with respect to the 4f spin, which changes $J$ from 4 to 3. Its gas-phase excitation energy of 207 mV[35] is comparable to our observed 163 mV conductance step (Fig. 1e,g), to which we assign this "6s-flip" excitation. The smaller excitation energy for these surface-adsorbed atoms compared to gas-phase ions is likely due to partial charge transfer out of the 6s orbital[28]. At high magnetic field, a gap in d$I$/d$V$ opens close to zero bias voltage (Fig. 1f) whose position gives the Zeeman energy. For an applied magnetic field $B = 6.5$ T this spin-flip excitation has an energy of 0.739 mV, yielding a g-factor $g = 1.97 \pm 0.01$ for both bridge-site orientations of Eu-b (Fig. S2), in approximate agreement with $g = 1.984$ reported for the lowest gas-phase Eu$^+$ multiplet[35]. Eu-b is thus a high-spin cation with vanishing orbital moment. Its spin $S = 4$ is larger than the largest previously observed ground-state spin ($S = 7/2$ for trivalent Gd) for any atom or ion of any element in a solid-state environment[36].

Using the g-factor for Eu determined by the tunneling spectra, we investigate ESR by using a magnetic field where spin-flip transitions are positioned in the highest accessible frequency range for our microscope, ~15–30 GHz. The ESR spectra of individual $^{153}$Eu-b and $^{151}$Eu-b atoms show an extremely rich peak structure (Fig. 2), consisting of several trees of resonant peaks. Each tree consists of six hyperfine peaks that correspond to the six projections $m_I$ of the nuclear spin $I = 5/2$ that is present in both isotopes. All previous ESR-STM spectra showing hyperfine peaks revealed only one such hyperfine-split tree[14,19,37] (Fig. S4). Here we detect at least four such trees for the following reasons: First, the large electron total angular momentum $J_{tot} = 4$ admits many electron spin transitions among its principal-axis projections $-4 \leq m_J \leq 4$. Second, the axial magneto-crystalline anisotropy $D$ is very small because of the absence of orbital angular momentum ($L = 0$ as a consequence of the exactly-half-filled f-shell), which places all the spin-flip transitions close in energy. Third, by choosing the polarity of the bias voltage $V_{DC}$, we are able to use spin torque[38] to selectively populate either the more positive or more negative values of $m_J$ as the initial state of ESR transitions (depicted schematically in Fig. 2e). A weighted sum of spectra taken with opposite bias voltages gives a spectrum spanning the full range from trees 1 through 8 (Fig. 3).

We find an excellent fit to the measured spectra for both isotopes and at both bias polarities by using the Hamiltonian

$$H = H_{Zee} + H_{tip} + H_{HF} + H_{CF} \tag{1}$$

where the electron Zeeman energy is $H_{Zee} = g\mu_B \mathbf{S} \cdot \mathbf{B}$; the magnetic interaction between tip and sample spins is $H_{tip} = g\, \mathbf{S} \cdot \mathbf{B}_{tip}$ where $\mathbf{B}_{tip}$ is the effective tip magnetic field; the dominant hyperfine (HF) interaction is $H_{HF} = A\, \mathbf{S} \cdot \mathbf{I}$; and the crystal field (CF) terms are

$$H_{CF} = D S_Z^2 + E(S_X^2 - S_Y^2) \tag{2}$$

Here $X$, $Y$, and $Z$ are the high-symmetry axes of the bridge binding site (Fig. 3a inset). We omit the weaker nuclear quadrupole interaction and higher-order CF terms.

The Eu$^+$ level scheme resulting from the Hamiltonian (Fig. 2e) shows the Zeeman splitting of the nine $m_J$ states by $B$ and $B_{tip}$. These nine electron spin states admit eight transitions compatible with selection rules that we take to require $\Delta m_J = \pm 1$ or $0$ and $\Delta m_I = 0$. We label these transitions as tree 1 ($m_J = -4 \leftrightarrow -3$) through tree 8 ($m_J = +3 \leftrightarrow +4$). Each $m_J$ state is split by the hyperfine interaction into six nuclear states $m_I$, which results in six ESR peaks in each tree. Because the Zeeman energy is much larger than both the hyperfine splitting and crystal field splitting, the quantum states are well approximated as product states of electronic and nuclear spin eigenstates, and we label them by their dominant $m_J$ and $m_I$ components. Considering only the HF and Zeeman effects, all eight electron spin transitions (trees) would have similar excitation energies. The CF term lifts this degeneracy by generating a blueshift for trees 1–4 and a redshift for trees 5–8. Comparing the spectra at opposite bias polarities shows that the highest-lying states are probed with negative bias and the lowest-lying ones with positive bias. For positive bias, spin torque drives the electron spin toward the ground state ($m_J = -4$), making tree 1 the most prominent in the spectra[38]. In marked contrast, at negative bias, the highest lying states (up to $m_J = +3$ and $+4$) are populated, giving tree 8 the largest intensity. The spin pumping is readily visible also in the tunneling spectra (Fig. S5), which provides an estimate of the intrinsic spin relaxation time $T_1 \approx 50$ ns[38].

The hyperfine coupling constant $A$ is given approximately by the hyperfine peak spacing seen in the spectra, which is ~300 MHz for Eu-153. Comparing the isotopes, $^{151}$Eu-b has more than twice the hyperfine splitting of $^{153}$Eu-b (Fig. 2). Fitting the Hamiltonian (Eq. 1) to the spectra yields hyperfine couplings $A = 291 \pm 13$ MHz for $^{153}$Eu-b, and $A = 670 \pm 13$ MHz for $^{151}$Eu-b (Fig. 3b). These hyperfine interactions are smaller by a factor of about two than those of the ions in the gas phase[39], likely because the s-shell spin density is reduced by delocalizing over neighboring surface atoms or by hybridization with the Eu 5d orbitals. Despite this reduced hyperfine magnitude, the measured ratio of $A$ for the isotopes on MgO ($A_{151} / A_{153} = 2.30$) agrees well with the ratio of 2.25 reported for gas-phase Eu$^+$ ions [39].

We determined the g-factor by taking the mean of two symmetrically positioned hyperfine peaks, which approximately eliminates the effects of the hyperfine and CF interactions. Extrapolation of this mean to zero current to eliminate most of the effect of tip fields (Fig. S7) yields a g-factor of ~1.97 for both isotopes, in good agreement with IETS measurements (Fig. S2).

To understand the effect of the magnetic anisotropy, we modeled the effect of orienting the CF axes corresponding to $D$ and $E$ (Eq. 2) along different symmetry axes of the bridge adsorption environment ($C_{2v}$ symmetry). We find that the measured spectra are consistent only with the anisotropy axis oriented along the surface normal, with $D \approx +1.55$ GHz, where the sign of $D$ indicates easy-plane (hard-axis) anisotropy (Fig. S8). A transverse term $E \approx +0.35$ GHz oriented along the Mg-Mg bridge direction significantly improves the fit to the data (Fig. 3b and Fig. S9). We note that similar CF parameters are typically found for Gd(III) ions in molecular spin labels, where $D \approx 0.5$–2 GHz and $E \approx 0.1$–0.4 GHz[40]. This similarity is reasonable in light of the $4f^7$ electron configuration of Gd(III), whose $f$ shell is isoelectronic with Eu$^+$. The vanishing orbital angular momentum in $4f^7$ ions has been demonstrated to suppress decoherence from spin-lattice relaxation[41]. Specially designed Gd(III) compounds show promise as candidate spin qubits [42,43] and spin *qudits*[36,44,45], which are $d$-dimensional ($d > 2$) quantum systems being explored as quantum computing platforms[46,47]. A crucial advantage of $4f^7$-based spin qudits is that the entire electron spin manifold can be accessed in typical ESR frequency ranges [45]. The Eu$^+$ studied here can be viewed as a $S = 4$ non-Kramers analogue to the Gd(III) $S = 7/2$ Kramers ion. The additional degrees of freedom due to the $I = 5/2$ nuclear spin of Eu result in a $d = 54$ combined electron-nucleus qudit manifold that may serve, like Gd(III), as a model high-dimensional single-ion qudit.

**Samarium spin excitations**

For bridge-site samarium (Sm-b), we observe two excitations in the tunneling spectra, at ~38 mV and ~148 mV (Fig. 4a,b). The magnetic field splitting of the 38 mV excitation indicates its magnetic origin (Fig. 4d and Fig. S11). Comparison to gas-phase spectra of Sm ions[35] yields an assignment of Sm-b to the singly-charged cation Sm$^+$ with ground-state configuration of $4f^6\,5d^0\,6s^1$ and electronic term $^8F_{1/2}$ ($S_{4f} = 3$, $L_{4f} = 3$, $J_{4f} = 0$, $S_{val} = 1/2$, $S_{tot} = 7/2$ and $J_{tot} = 1/2$). It differs from Eu$^+$ in having one fewer $f$ electron, which endows it with a large orbital angular momentum $L = 3$. The first excitation of the free ion is at 40.5 mV, a transition to $^8F_{3/2}$ that excites the $f$-shell from $J_{4f} = 0$ to 1 by "tilting" $L$ with respect to $S$, to which we assign the observed 38 mV transition (Fig. 4e). The next higher excitation of the free ion that is compatible with inelastic tunneling selection rules ($\Delta m_J = \pm 1$ or 0) occurs at 188 mV and excites from the ground state to the $^6F_{1/2}$ multiplet, where $S_{4f}$ and $S_{val}$ become antiparallel in a $6s$-flip similar to Eu$^+$ (Fig. 4e). For the surface adsorbed species Sm-b we detect this transition at ~148 mV, which is similar to the gas-phase value, but smaller for the reasons discussed for Eu$^+$. For oxygen-site Sm (Sm-O), the "$L$-$S$-tilt" excitation splits into two $J = 3/2$ doublets, at 26.2 and 88.8 mV, because of strong magneto-crystalline anisotropy (Fig. S12). In a magnetic field, Sm-b exhibits a low-energy spin-flip excitation, at ~1.8 mV for $B = 6.5$ T (Fig. 4c). This excitation occurs at energies that correspond to $g = 4.75 \pm 0.01$ for the horizontally oriented bridge site (Sm-bh) and $g = 5.03 \pm 0.01$ for the vertically oriented bridge site (Sm-bv). For Sm-O, it occurs at 1.55 mV at 6 T, giving $g = 4.46 \pm 0.01$ (Fig. S12). These $g$-factors are notably higher than both the Sm$^+$ gas-phase ion, for which $g = 3.950$ [35], and the calculated Landé $g$-factor of 4 (given $L = 3$ and $S = 7/2$). These $g$-factors are extraordinarily high for atomic spins, and they arise from the large but nearly cancelling orbital and spin angular momenta within the atom (Figs. S13a, S14).

Sm and Eu are the first lanthanide atoms known to exhibit inelastic tunneling excitations when adsorbed on MgO/Ag(100), in contrast to the featureless spectra seen previously for Dy, Ho, and Er on this surface[3,24,32]. We note that these tunneling spectra closely resemble the spectra reported for Sm and Eu on graphene on Ir(111)[28]. When adsorbed on the Mg sites of MgO, in contrast, both Sm and Eu exhibit featureless d$I$/d$V$ spectra (Figs. S2F and S12), which we propose is the result of a change in charge state compared to the bridge and O sites, and consequent pairing of the valence-shell electrons.

The most abundant Sm nuclear spin is $I = 0$ for five isotopes, which show a single ESR peak (Fig. 5a, top). The two remaining isotopes have $I = 7/2$ so they display eight hyperfine peaks (Fig. 5a, bottom). From fitting these spectra, we derive $A = 170 \pm 6$ MHz for $^{149}$Sm-b and $A = 205 \pm 8$ MHz for $^{147}$Sm-b. The ratio of these two hyperfine couplings (1.21) matches the ratio of their nuclear magnetic moments ($-0.8035 / -0.6631 = 1.21$) [48], suggesting that the different nuclear moments are the origin of the different hyperfine parameters. Magnetic resonance images of individual Sm atoms each show concentric resonant slice rings, one for each nuclear spin state (Fig. 5b–d and Fig. S15), which provides visualization of the nuclear spin states and the tip-adatom magnetic interaction[49].

We expect the magneto-crystalline anisotropy for adsorbed Sm$^+$ to be large because of its large orbital angular momentum $L = 3$, as exhibited by the large splitting of the $L$-$S$-tilt excitation in Sm-O (Fig. S12). The ground-state multiplet of Sm is a Kramers doublet, having $m_J = \pm 1/2$, so it cannot exhibit zero-field splitting. It can, however, exhibit anisotropy in the $g$-factor. Comparing the two Sm bridge orientations, ESR spectra yield effective $g$-factors of $4.71 \pm 0.05$ and $5.04 \pm 0.08$ for Sm-bh and Sm-bv, respectively (Fig. S14), in agreement with the values obtained above from high-field tunneling spectra. This pronounced $g$-factor difference between sites implies a large in-plane anisotropy of the $g$-tensor, considering that our magnetic field is applied ~42° away from the O-O axis of Sm-bh, which differs only slightly from the ~48° angle to the Sm-bv axis (Fig. 1d). This large in-plane $g$-tensor anisotropy may be useful for direction-dependent magnetic field sensing using Sm-b as a single-atom magnetometer.

## Conclusions

The Eu$^+$ and Sm$^+$ studied here constitute an unusual case of stable, open-shell lanthanide cations in a solid-state system. The key ingredients for realizing the open-shell state in this work include the greater propensity for Eu and Sm to be divalent in compounds, compared to other lanthanides[30], and the electron transfer from each atom to the metal substrate through the thin insulating film. This electron transfer is well-known for adsorbates on metal-supported MgO thin films[21,22,50,51] and a variety of other insulating films on metal substrates[52,53]. We therefore propose such adsorption and charging as a general strategy for stabilizing and studying uncommon monovalent lanthanide species (see Supplementary text section 8).

Our demonstration of ESR-STM on monovalent lanthanide atoms opens electronic access to these versatile elements one atom at a time and allows the determination of their magnetic quantum properties as a function of local atomic environment. With a single ESR spectrum, we determine the isotope, hyperfine interactions, and magneto-crystalline anisotropy. We anticipate that the relatively large hyperfine coupling in these atoms will facilitate direct control of their nuclear spins[19] with potentially much longer lifetimes than for electron spins[4]. The use of recently demonstrated remote driving methods[54] will additionally lengthen relaxation times, allowing qudit gate operation.

# Methods

**Microscope and sample.** Measurements were performed in a home-built ultrahigh-vacuum STM operating at 0.6–1.2 K as noted. An MgO thin film was grown on an atomically clean Ag(001) single crystal held at ~340 °C, by thermally evaporating Mg from a Knudsen cell in a $1\times10^{-6}$ torr $O_2$ environment at ~0.5 monolayer (ML) per minute for an average coverage of ~1.6 ML MgO. This produced two-ML MgO terraces interspersed with clean Ag(001) regions used for tip preparation. Metal atoms of Fe, Ti, Eu, and Sm were deposited one element at a time by e-beam evaporation from pure metal pieces onto the sample held at ~10 K. Previous studies indicated that the Ti atoms may be hydrogenated (TiH)[14–18,20] and we describe them here as Ti for simplicity. An external magnetic field was applied at an angle ~9° out of the surface plane with the in-plane component aligned nearly along the [100] direction of the MgO lattice, as described in the text. STM images were acquired in constant-current mode and all voltages refer to the sample voltage with respect to the tip. Voltages and energies are treated as interchangeable, where the factor of the elementary charge e is implied when giving energies in units of mV.

**ESR apparatus.** The ESR spectra were acquired by sweeping the frequency of an RF voltage generated by an RF generator (Agilent E8257D) across the tunneling junction and monitoring changes in the tunneling current. The signal was modulated at 337.11 Hz by chopping VRF, which allowed readout of the spectrum from the measured tunnel current (room-temperature electrometer model FEMTO DLPCA-200) applied to a lock-in amplifier (Stanford Research Systems SR830). The RF and DC bias voltages were combined at room temperature using an RF diplexer and carried to the STM tip through semi-rigid coaxial cables with a loss of ~30 dB at 20 GHz.

**STM tips.** The STM tip was a mechanically sharpened bulk iridium wire that was presumably coated with silver by indentation into the sample. It was prepared in vacuum by field emission and by indentations into the Ag sample until the tip gave a good lateral resolution in STM images. All tips used the same bulk Ir wire but differed in the elemental identity and precise arrangement of atoms near the apex. To prepare a spin-polarized tip, about 2–8 Fe atoms were each transferred from the MgO onto the tip by applying a bias voltage (~0.55 V) while withdrawing the tip from near point contact with the Fe atom. The ESR spectra shown in the main text were recorded with "Tip 1", an STM tip apex that yielded unusually high ESR contrast on $^{48}$Ti-b atoms, where $\Delta I/I$ was about 9% as shown in Fig. S5, compared to ~2–4% observed for typical spin-polarized tips used for ESR. This tip produced spin-torque in the Eu-b electron spin, and showed strong bias-voltage dependence for tunneling spectra of Ti-O, which indicates spin polarization (Fig. S5). It was obtained as for other ESR tips by picking up several Fe atoms onto a presumably Ag-coated tip. To illustrate the signal-to-noise given with this particular tip, it was used to obtain ESR spectra of bridge-site Ti on this surface (Fig. S4). All ESR data were acquired using Tip 1 except where noted. It was used to acquire all ESR data shown in the main text Figs. 2, 3 and 5, and Supplementary Figs. S3, S4, S6, S7, S14d–g, and S15.

**Atom manipulation.** Eu and Sm atoms were repositioned on the surface by pulling them with the STM tip while close to point-contact, by using bias $V_{DC}$ = 10 mV and tunnel current $I$ = 2–6 nA. We observed that it was always possible to position the atoms at Mg sites, but most tips tended to give either O or bridge sites preferentially, with bridge-favoring tips being much more frequent. We speculate that the different binding site preferences for manipulation using different tips may arise from different local work functions (tunneling barrier heights) of each tip apex material and atomic arrangement, and consequently of different electric fields present in the junction, which can readily dominate the bias-voltage dependence of the electric field. Manipulation and binding-site examples are shown in Fig. S1.

## Acknowledgments

We are grateful to Stefano Rusponi for helpful discussions and to Bruce Melior for his outstanding technical support. CPL and GC acknowledge support from the Department of the Navy, Office of Naval Research under ONR award number N00014-21-1-2467. HB acknowledges funding from Swiss National Science Foundation Advanced Grant TMAG-2_209266.

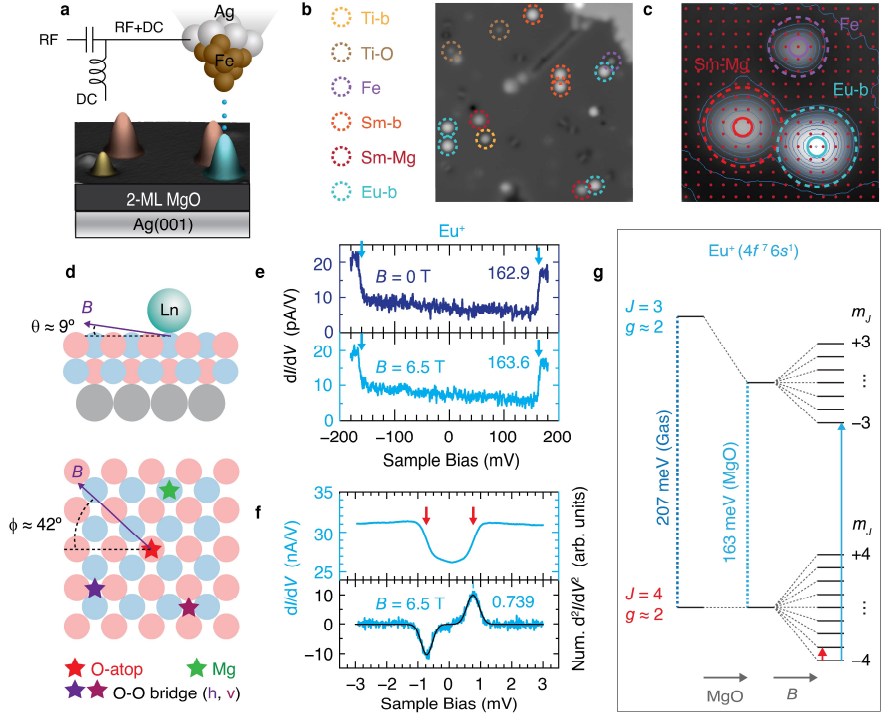

**Fig 1 | Lanthanide atoms on an MgO film and Eu excitations. a**, Schematic of the STM tip and sample consisting of adatoms on a two-monolayer epitaxial MgO film on Ag(001). **b**, Constant-current STM image of adatoms as labeled (tip-height setpoint $I_{set}$ = 20 pA, $V_{set}$ = 60 mV, 24 nm square). Atoms were identified by their spectral features and their adsorption sites. **c**, STM image used to determine binding sites (red points denote O atoms), setpoint 20 pA at 60 mV, 4.6 nm square. **d**, Side and top views of MgO structure (red: O; blue: Mg; grey: Ag) showing lanthanide binding sites (star symbols as labeled) and applied magnetic field direction. **e,f**, Inelastic tunneling spectra (d$I$/d$V$) of bridge-site Eu adatoms (Eu-b). **e** shows excitations at ~163 mV at 0 T and 6.5 T with the tip positioned 0.1 nm farther away from the surface from tip height $I_{set}$ = 4 pA at $V_{set}$ = 80 mV. $V_{mod}$ = 3 mV r.m.s. (root mean square) at $f$ = 803 Hz. **f** shows low-energy excitation appearing in a magnetic field $B$ = 6.5 T, setpoint 300 pA at 10 mV, $V_{mod}$ = 0.1 mV r.m.s. The lower curve is d$^2I$/d$V^2$, numerically computed from the upper curve. **g**, Schematic energy level diagram of gas-phase and MgO-adsorbed Eu$^+$ ions. Application of a magnetic field $B$ (right) splits the states (not shown to scale). Vertical arrows show the two types of observed inelastic excitations: $m_J$-changing (red) and 6$s$-flip (blue).

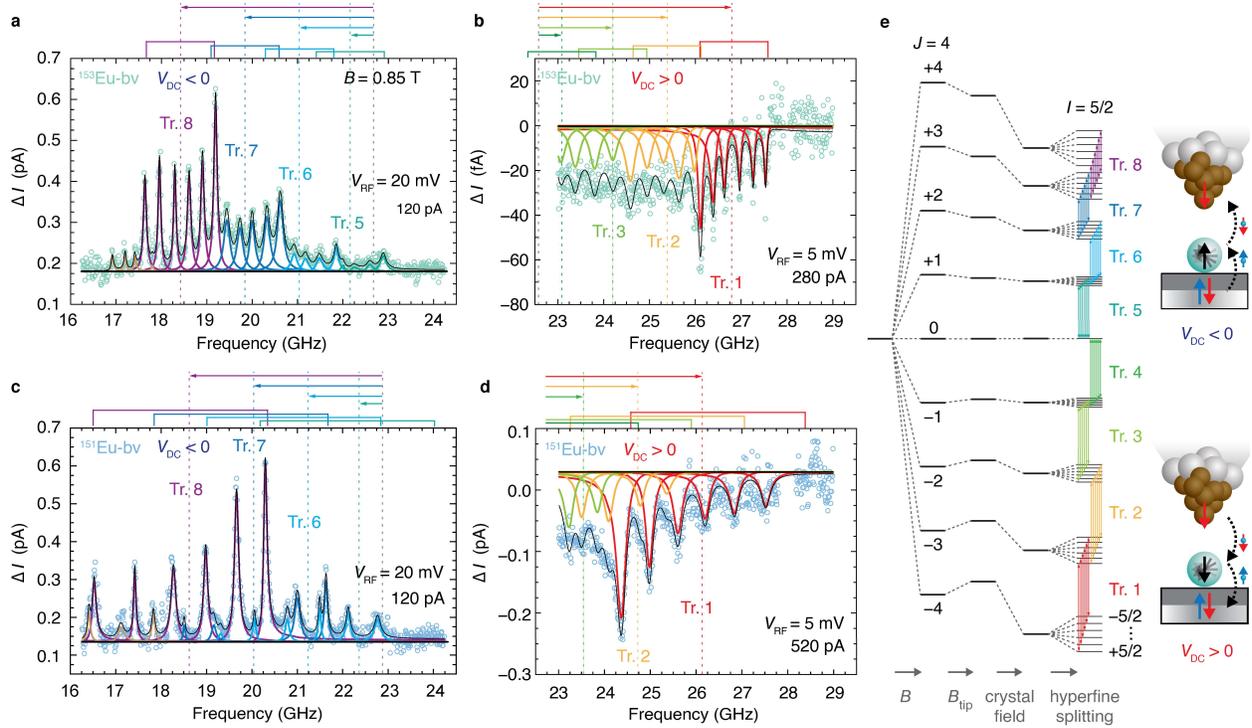

**Fig. 2 | ESR spectra of Eu isotopes on MgO bridge sites and magnetic level scheme. a,b**, $^{153}$Eu-b (nuclear spin $I = 5/2$, natural abundance 52.19%) at negative (**a**) and positive (**b**) bias voltage $V_{DC}$. Color code indicates tree number (initial-state $m_J$) and colored arrows show shift of each tree away from the Zeeman energy due to the out-of-plane magneto-crystalline anisotropy $D$. **c,d**, $^{151}$Eu-b ($I = 5/2$, 47.81%) at negative and positive bias. All spectra: setpoint $V_{set} = 50$ mV, $I_{set}$ and radio-frequency voltage $V_{RF}$ (zero-to-peak) as shown in each panel, $T = 1.2$ K, $B = 0.85$ T; bias $V_{DC} = -50$ mV for **a** and **c**, $V_{DC} = +50$ mV for **b** and **d**. Data points in **a**–**d** have been omitted in frequency intervals that are dominated by RF transmission artifacts; see Fig. S3 for unmodified data sets. **e**, Schematic magnetic energy levels of the ground-state multiplet, split by Zeeman energy due to $B$ and $B_{tip}$, perturbed by $D > 0$ and hyperfine splitting with $A > 0$ (shifts are exaggerated), to yield eight trees of transitions (colored lines) labeled Tr. 1 through Tr 8. The rightmost schematics depict the electron spin-torque process at positive bias (bottom) and negative bias (top).

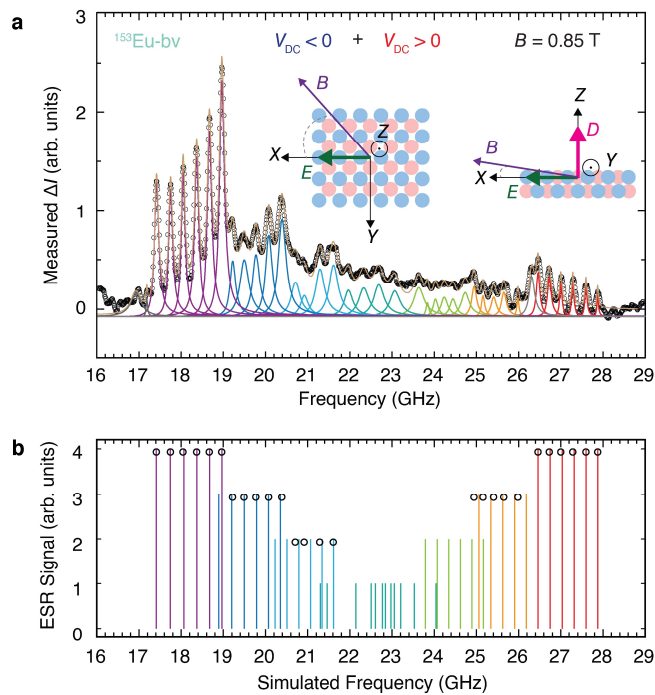

**Fig. 3 | Broad-band ESR spectrum of Eu and Hamiltonian model transition energies. a**, ESR spectrum of bridge-site $^{153}$Eu over a wide frequency range that covers all eight hyperfine-split trees. Spectrum shown is the weighted sum of two spectra, one at each bias polarity (see Fig. S6 for details), which renders the transitions at each extreme of the spin-flip energy range visible. **b**, Frequencies of allowed transitions determined from the model Hamiltonian with fit parameters $B_{tot} = B + B_{tip} = 0.8175$ T at angle 84° from the surface normal and 40° from MgO bridge-site O-O direction, $A = 291$ MHz, crystal field parameters $D = 1515$ MHz and $E = 360$ MHz along axes shown in **a** inset. Each transition is shown as a vertical line having height and color that indicates the dominant $m_J$ component of the initial state but are otherwise arbitrary. Black circles represent fitted peak positions from the data in **a**, with fits for trees 3–5 omitted due to large uncertainties in their peak positions. The spacing of peaks in trees 4 and 5 are irregular because the initial or final state has $m_J = 0$, for which the electron Zeeman energy vanishes, so other terms determine the properties.

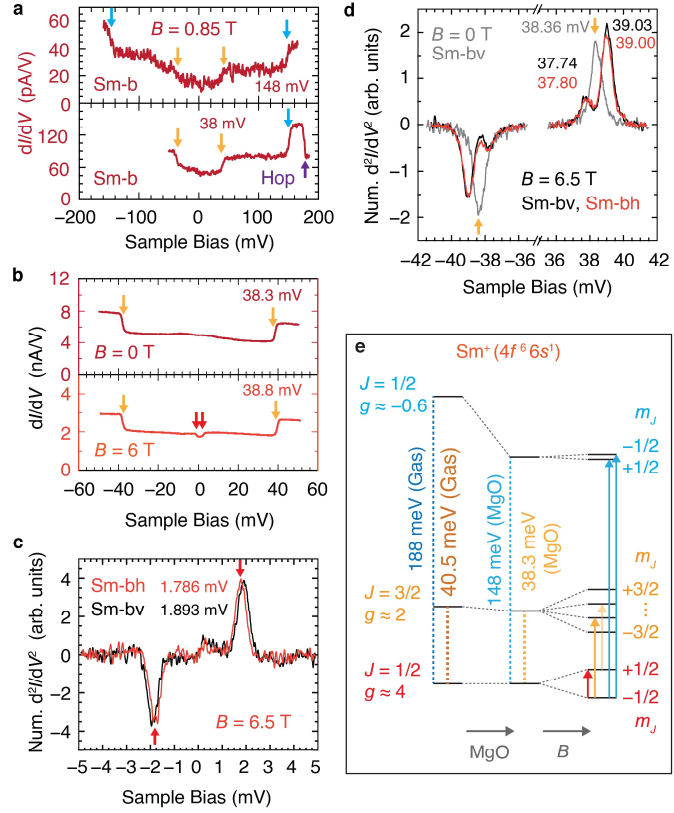

**Fig. 4 | Excitations of samarium. a,** Tunneling spectra (d$I$/d$V$) of bridge-site Sm on MgO (Sm-b). Conductance steps show excitations at ~148 and at ~38 mV. Top: low conductance (tip-height setpoint $I_{set}$ = 2.5 pA at $V_{set}$ = 80 mV); bottom: higher conductance (5 pA at 80 mV), where the Sm atom hops as indicated due to larger voltage and current, limiting the accessible voltage range. **b,** Spectra of Sm-b showing ~38 mV step at 0 T and at 6 T. The excitation shifts with magnetic field, and a low-energy excitation is visible at 6 T. **c,** Low-energy tunneling spectra of bridge-site Sm showing Zeeman splitting of the ground state doublet. Energies as labeled (uncertainty 0.004 mV) for Sm-bh (red) and Sm-bv (black). Tip-height setpoint 200 pA at 10 mV, $V_{mod}$ = 0.2 mV rms. **d,** Tunneling spectra near the ~38 mV excitation of Sm-b with energies in mV as labeled. Spectra are at $B$ = 0 (grey) and at $B$ = 6.5 T, black for Sm-bv and red for Sm-bh. At 6.5 T the excitation splits into peaks of different amplitudes. The lower energy (and lower amplitude) peak is assigned to a hot band (excitation from the $m_J$ = +1/2 heated state, Figs. S11 and S13). Setpoint 50 pA at 10 mV, $V_{mod}$ = 0.2 mV rms. Effect of the bridge orientation (Sm-bv versus Sm-bh) is negligible. Spectra in **c** and **d** are d$^2I$/d$V^2$ obtained numerically from measured d$I$/d$V$ (Fig. S11). **e,** Schematic energy level diagram of gas-phase and MgO-adsorbed Sm (Sm-b). Application of a magnetic field $B$ (right) splits the states. Vertical arrows show transitions assigned in the measured spectra of Sm-b: $m_J$-changing (red), $L$-$S$-tilt from the ground state (orange) and from an excited state (yellow), and 6$s$-flip (blue).

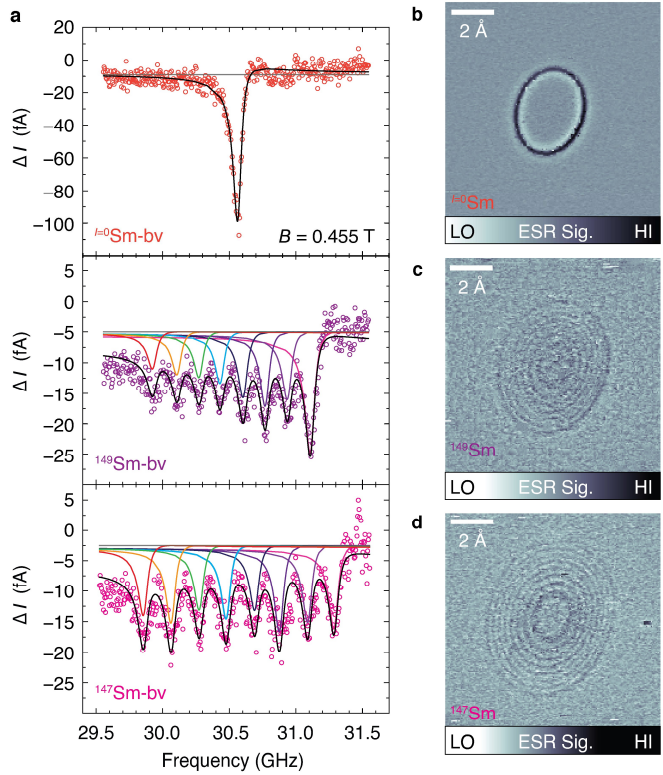

**Fig. 5. ESR spectra and magnetic resonance images of Sm isotopes. a**, ESR spectra of different Sm isotopes: Zero-nuclear-spin isotope of Sm-b ($I = 0$, 71.19% total for $^{144,148,150,152,154}$Sm), $^{149}$Sm-b ($I = 7/2$, 13.82%), and $^{147}$Sm-b ($I = 7/2$, 14.99%). All spectra: tip-height set-point 20 pA at 25 mV, $V_{RF}$ = 5 mV, $T$ = 0.6 K, $B$ = 0.455 T, Sm at a vertical bridge site of MgO (Sm-bv). **b–d**, Magnetic resonance images show concentric rings where the tip field $B_{tip}$ shifts the atom into resonance at $f_{RF}$ = 29.0 GHz. The $I = 0$ and $I = 7/2$ isotopes show one and eight rings, respectively. Setpoint 80 pA at 25 mV, $V_{DC}$ = −15 mV, $V_{RF}$ = 20 mV.

# Supplementary Information

## Contents

# Supplementary Notes

## 1. Eu and Sm adsorption site stability

After deposition onto the cold MgO surface, both Eu and Sm could be found on either the Mg-site or bridge site (b-site). However, it was much more common to find Eu on the b-site and Sm on the Mg-site. In addition, it was found to be significantly easier to cause the Sm atoms to hop from a b-site to an Mg-site compared to Eu atoms. Either close tip proximity or elevated bias (or both) caused the atoms to hop to the Mg-site, and the voltage and proximity thresholds for this hopping were consistently lower for Sm-b than for Eu-b. This hopping threshold additionally showed significant variation among tip apexes. It was much more difficult to re-position Sm or Eu to the O-site than to the b-site. The ability to position Sm or Eu on the O-site depended strongly on the particular tip apex used, with certain rare tip apexes capable of repositioning Sm to the O-site reproducibly. We encountered only one tip apex that was consistently able to reposition Eu to the O-site. This may imply that both atoms have a shallow barrier to diffusion from the O-site.

## 2. Determination of ground state g-factors

The *g*-factors were determined for Eu-b and for Sm-b by two methods: by using IETS with a non-spin polarized tip, and by ESR spectra using a magnetic tip.

Although the magnetic anisotropy for Eu-b is expected to be very small because of its lack of orbital angular momentum, to explore the possibility of *g*-factor anisotropy we acquired IETS spectra for both Eu-bh and Eu-bv (Fig. S2a,b). The measured Zeeman energies at $B = 6.5$ T for the two species are identical within the uncertainty, such that we resolve no *g*-factor anisotropy with our external field applied ~42° with respect to the h-bridge direction (~48° to the v-bridge direction). The Zeeman energy of both Eu-b orientations implies a *g*-factor 1.966 (mean of 1.967 for Eu-bh and 1.964 for Eu-bv), close to the $Eu^+$ gas phase value 1.984.

For Sm-b, a similar comparison of the 6.5 T tunneling spectra between the two orientations, Sm-bh and Sm-bv, using a non-spin polarized tip is shown in main text Fig. 4c and Fig. S11a. In contrast to Eu-b, Sm-b shows significant $g$-factor anisotropy as evidenced by the $> 0.1$ mV difference in the Zeeman splitting between the two orientations. These spectra give apparent $g$-factors of 4.747 for Sm-bh and 5.031 for Sm-bv for this applied direction for $B$.

To obtain reliable $g$-factors from ESR, we must remove the effect of the tip magnetic field to the extent possible on the measurement since this magnetic field is unknown and uncharacterized for each tip apex. We compensate for the effect of the exchange interaction with the tip by extrapolating tip-height-dependent peak energies to zero tunnel current (Figs. S7 and S14 for Eu-b and Sm-b, respectively) such that only the externally applied magnetic field $B$ is expected to act on the atom, and the effect of $B_{tip}$ is expected to be largely eliminated or reduced to a long-range magnetic dipole component. For Eu-b in particular, in order to remove in addition the effect of the CF we take the average energies of ESR peaks that have opposite-sign CF shift and opposite-sign hyperfine shifts. These are found by identifying the symmetrically opposite hyperfine peaks in trees 1 and 8, which are found at opposite bias (Fig. S7c,f). Since the bias voltages are chosen symmetrically around zero bias, this may also remove small but systematically observed Stark shifts caused by the bias voltage. For $^{151}$Eu-bh, this yields $E_{Zee}$ = 23.52 ± 0.05 GHz at 0.85 T, or $g$ = 1.977 ± 0.004. For $^{153}$Eu-bv we find $E_{Zee}$ = 23.36 ± 0.05 GHz at 0.85 T, which yields a nearly identical value of $g$ = 1.964 ± 0.004. Both values are close to that obtained from IETS spectra (1.966) and to the gas phase value for Eu$^+$ (1.984). A similar comparison, for Sm-bh and Sm-bv, made with the same tip apex is shown in Fig. S14d–g.

For a given tip, the $g$-factors of both elements measured by ESR were found to be systematically either higher or lower than the $g$-factors derived from tunneling spectra (conductance steps measured using non-spin-polarized tips). The ESR measurements require a spin-polarized tip, making exchange coupling to the tip and dipole magnetic fields from the tip unavoidable. The distance-dependent spectra show that the dominant tip-sample interaction is linear with tunnel current, and therefore of exchange character. Extrapolating to zero tunnel current removes the effect of exchange interaction, but the small dipolar magnetic field of the tip is not removed by extrapolating to zero current because it decays too slowly as the current is reduced. The tip magnetic dipole moments are expected to vary significantly among different tip apexes owing to the random, uncontrolled nature of each Fe tip preparation, which may explain the significant tip-dependent variations in Zeeman energy found for Sm-bh and Sm-bv species. This stray tip magnetic field is additive with the external magnetic field. The effects of this uncompensated dipole field are illustrated by the variation among Sm-bh and Sm-bv $g$-factors derived from ESR using different tips as shown in Fig. S14. The lateral tip position can also influence the tip field sensed by the atom, despite extrapolating to the limit of zero tunneling current, as shown in Fig. S14a. Nevertheless, Sm-bv was always found to have a higher resonant frequency than Sm-bh, as for Ti-bv compared to Ti-bh. Overall, despite variations in tips, the apparent $g$-factors for Sm-bh and Sm-bv obtained from ESR measurements (Fig. S14) differed by no more than 2% from the values obtained from IETS (Fig. 4c and Fig. S11a).

Interestingly, we also observed strong ESR signal intensity variation between Sm-bh and Sm-bv, which depended on the tip apex. "Tip 1" and "Tip 3" data showed stronger signal intensity for Sm-bv compared to Sm-bh (for "Tip 3" the difference was a factor of 5 in amplitude), while "Tip 2" and "Tip 4" instead showed stronger signal intensity for Sm-bh. Signal intensity differences between Ti-bh and Ti-bv could also be observed, but the relative change in intensity between the two species of Ti was much less compared to the difference for Sm. We note that in the data shown here, the intensity variations between Ti and Sm qualitatively follow each other (*i.e.* when Sm-bh has a more intense signal than Sm-bv, Ti-bh similarly has a more intense signal than Ti-bv, and vice-versa). We speculate that this asymmetry may be related to the in-plane components of the tip dipole field, which may provide greater (or lesser) driving strength for one bridge orientation or the other due to the in-plane $g$-factor anisotropy of Sm-b and Ti-b.

We observed that with most tips, the ESR signal for Sm-b appeared as a "dip" (decrease in conductance) at positive bias, similar to Eu-b, while ESR signals for Ti-b and Ti-O appeared as "peaks" (increase in conductance) under similar conditions when the same tip apex was used. This consistent observation suggests opposite-sign tunnel magnetoresistance for the lanthanides compared to Ti.

To determine the g-factor of the Eu-b multiplet at ~163 mV (the 6s-flip excitation) we analyze the measured shift in the excitation as the field is changed from 0 to 6.5 T, which is $0.7 \pm 0.1$ mV (Fig. 1e). We model the shift as the difference in initial and final Zeeman energies $g_f \mu_B B m_f - g_i \mu_B B m_i$, where $m_i = -4$, $m_f = -3$ (the only transition from the ground state allowed by $\Delta m_J = \pm 1$ or 0), $g_i = 1.97 \pm 0.01$ is the ground multiplet g-factor, and $\mu_B B = 0.3762$ at 6.5 T. This yields an excited state g-factor $g_f = 2.0 \pm 0.1$, which is consistent with the gas-phase Eu$^+$ ion's value of 1.981 for this excited state.

## 3. Modeling Eu magnetic anisotropy

To understand the effect of the magnetic anisotropy on Eu-b expressed by the crystal field terms in the Hamiltonian (main text Eqs. 1 and 2, illustrated in Figs. S8 and S9), we first consider the effect of a hypothetical purely axial crystal field $DS_z^2$ aligned to the magnetic field. This anisotropy would result in a quadratic dependence of the energy on $S_z$, and consequently a linear dependence of the ESR frequency on the initial-state quantum number $m_J$, as observed to good approximation in the spectra and depicted in Fig. S8b. However, the field direction is not aligned to any of the symmetry axes of the MgO bridge site in our experiment (main text Fig. 1d). Assigning the Hamiltonian principal anisotropy axis to an in-plane symmetry direction $X$ or $Y$ yields a very poor fit to the spectra (Fig. S8d), ruling out this possibility. In contrast, an out-of-plane hard-axis anisotropy yields the nearly linear spacing observed (Fig. S8c). Assignment of hard-axis anisotropy in the out-of-plane direction is the only assignment that satisfies the known geometric symmetry and peak-height ordering that we observe. A qualitative fit yields an out-of-plane crystal field parameter of $D \approx 1.5$ GHz $\approx 6.2$ $\mu$eV. This crystal field parameter is very small compared to other surface-adsorbed lanthanides[32], a consequence of Eu having zero orbital angular momentum. Simulated spectra shown in Fig. S8b–d neglect the hyperfine coupling for simplicity, while the spectra shown in Fig. S8e uses hyperfine coupling $A = 700$ MHz, which is qualitatively consistent with $^{151}$Eu. One can then compare Fig. S8e with the $^{151}$Eu spectra shown in main text Fig. 2c,d. The corresponding energy eigenvalues are shown in Fig. S8f while excitation energies are shown in Fig. S8g, both as a function of $B$. At low $B$, the magnetic anisotropy $D$, Zeeman energy and hyperfine coupling $A$ all become comparable in energy scale, resulting in complex changes to the excitation spectrum.

Since Eu-b binds to the bridge site with $C_{2v}$ symmetry, transverse magnetic anisotropy is also expected. This situation is depicted schematically in Fig. S9a. The inclusion of the transverse anisotropy parameter $E$ changes the apparent separation in excitation energies between trees (different electron spin excitations). Simulated spectra for a Hamiltonian assuming transverse anisotropy $E = 0$ are shown in Fig. S9b, and the inclusion of finite $E$ close to the upper allowed value $E = \frac{1}{3}D$ is shown in Fig. S9c,d for two different orientations of external magnetic field relative to the transverse anisotropy axis $E$. A key indicator of the presence of transverse anisotropy is its effect on the apparent crystal field splitting (spacing between trees) for Eu-bh compared to Eu-bv. Depending on whether $E$ is oriented along the O-O or Mg-Mg direction, following Fig. S9c,d, numerical modeling indicates that one orientation of Eu-b has an overall larger apparent spacing between trees and the other Eu-b orientation has overall smaller spacing. This transverse anisotropy also has the effect of expanding the excitation energy spacing between trees 1 and 2 relative to trees 7 and 8, as compared to the case where no transverse anisotropy is present. ESR measurements confirm these adsorption-site dependencies on the tree spacing as shown in Fig. S9e–i. We find the transverse parameter $E \approx 350$ MHz = 1.45 $\mu$eV.

## 4. Nature of the splitting of the 38 mV excitation of Sm-b

This section provides evidence that the splitting of the ~38 mV IETS peak seen for Sm-b at high magnetic field $B$ arises from heating of the field-split ground-state multiplet, not from transitions from the ground state to multiple final states. Consequently, only a single transition is observed when starting from a given state ($m_J = \pm\frac{1}{2}$) in the ground multiplet ($J = \frac{1}{2}$).

As shown in main text Fig. 4d and in Fig. S11b,f,h, the 38.4 mV excitation splits at high magnetic field $B = 6.5$ T into a strong 39.0 mV peak and a weaker 37.8 mV peak (values are averages from Fig. 4d). These high-field peaks are spaced symmetrically about the 38.4 mV zero-field peak. Based on the similarity of the energy to the 40.5 mV free Sm+ excitation, we assign the peaks to different allowed "L-S-tilt" excitations, in which $J = \frac{1}{2} \to J = \frac{3}{2}$. The intensity of the smaller peak varied dramatically as a function of tunnel current setpoint used during data acquisition (Fig. S11c). Although it is plausible that IETS selection rules vary depending on the tip proximity to the f-shell electrons, we propose that the much more likely explanation is that the tunnel current heats the Sm-b spin from the magnetic ground state to populate both of the $m_J = \pm\frac{1}{2}$ states through random IETS spin-flip events. The Zeeman energy required to flip the Sm-b from the ground state to the first excited state, $(m_J = -\frac{1}{2} \to +\frac{1}{2})$, is much lower in energy than the 38 mV L-S-tilt excitation. For that reason, use of bias voltages near 38 mV to observe that transition randomize the Sm-b ground multiplet through spin-flip IETS events to equally populate $m_J = -\frac{1}{2}$ and $m_J = +\frac{1}{2}$, in the limit of large current. The precise ratio depends on the tunnel current, the lifetime of the $m_J = +\frac{1}{2}$ excited state, and the tip spin polarization, which we believe to be near zero. As the time-average population of $m_J = +\frac{1}{2}$ increases, L-S-tilt excitations increasingly correspond to $\left|\frac{1}{2}, +\frac{1}{2}\right\rangle \to \left|\frac{3}{2}, +\frac{1}{2}\right\rangle$, and reduce the dominant $\left|\frac{1}{2}, -\frac{1}{2}\right\rangle \to \left|\frac{3}{2}, -\frac{1}{2}\right\rangle$ transitions from the ground state. Consequently, as the tunneling current increases, the intensity of the 37.8 mV transition increases, so that the ratio of the 39.0 mV IETS signal to the 37.8 mV IETS signal decreases. This ratio is plotted in Fig. S11d and shows an exponential dependence on the tunneling current. We cannot at present explain why the ratio appears to converge to ~1.5 rather than 1 as expected for an equally-randomized ground multiplet population expected for a non-spin-polarized tip[38]. (A similar weaker, lower energy peak was observed for the 26.2 mV excitation of Sm-O as seen in Fig. S12g–j). From these observations we conclude that the 37.8 mV excitation arises from tunneling electrons heating the ground state. Consequently, there is only one observed L-S-tilt transition for each initial state ($J = \frac{1}{2}$ to $\frac{3}{2}$), and the two observed peaks arise from occupation of two initial states in the ground multiplet.

## 5. State assignments for the 38 mV excitation of Sm-b

Here we consider alternative interpretations of the observed excitations seen in tunneling spectra at ~38 mV for Sm-b, which is the excitation energy required to tilt $L$ with respect to $S$ that we term the "$L$-$S$ tilt" excitation. When starting from the $J = 1/2$, $m_J = -1/2$ ground state, one might expect transitions to any of the four states in the $J = 3/2$ excited state multiplet, subject to any selection rules for such transitions. We consider how the measured excitation energy shifts in a magnetic field, and interpret the g-factors implied by this shift for each case in order identify the likeliest value of $m_J$ for the final state.

In the following we make use of the measured excitation energies $\Delta E_0 = 38.36$ mV measured by IETS at $B = 0$ T, and a field-shifted energy $\Delta E_{6.5\,T} = 39.03$ mV measured at $B = 6.5$ T (Figs. 4D, S11b). This gives a net Zeeman shift $\Delta E_{6.5\,T} - \Delta E_0 = 0.67$ mV. (We disregard the weaker peak that shifts in a magnetic field by an equal amount downward, to 37.74 mV (Figs. 4D, S11b), by attributing it to the excitation from the $\left|\frac{1}{2}, +\frac{1}{2}\right\rangle$ state. This initial state is occupied only due to heating of the ground-state multiplet, as discussed in section 4.)

**Small anisotropy case**. We first consider the case where $D$ is significantly smaller than the Zeeman energy at large magnetic field, in which case the $J = 3/2$ spin is oriented by the external magnetic field like a free spin. In this situation, the excitation energy for a transition $|J_i, m_i\rangle \rightarrow |J_f, m_f\rangle$ is given by the sum of the zero-field excitation energy $\Delta E_0$ and the change in Zeeman energy:

$$\Delta E = E_f - E_i = \Delta E_0 + g_f \mu_B B m_f - g_i \mu_B B m_i \qquad (\text{Eq. S1})$$

where $g_i$ and $g_f$ are the g-factors for the initial and final states, and $\mu_B$ is the Bohr magneton. These g-factors are effective g-factors under the conditions of our applied field direction, which is not along any of the high-symmetry directions of the binding site, so we expect that they may fall somewhere in the range spanned by the g-factors for each axis in the presence of the crystal field.

Since we lack measurements in a vector magnetic field, knowledge of the axis-dependent g-factors is beyond the scope of this analysis. Solving for the final state g-factor gives

$$g_f = \frac{\Delta E - \Delta E_0 + g_i \mu_B B m_i}{\mu_B B m_f} \quad \text{(Eq. S2)}$$

We consider transitions $\left|\frac{1}{2}, -\frac{1}{2}\right\rangle \to \left|\frac{3}{2}, m_f\right\rangle$ for each possible value of $m_f = -\frac{3}{2}, -\frac{1}{2}, +\frac{1}{2},$ and $+\frac{3}{2}$. Using $\mu_B B = 0.2376$ mV at $B = 6.5$ T, and the measured initial-state g-factor $g_i = g_{\frac{3}{2}} = 5.03$ (Fig. 4 and Fig. S11, choosing values that apply to Sm-bv), Eq. S2 yields respective final-state g-factors $g_{\frac{3}{2}} = 0.49, 1.47, -1.47,$ and $-0.49$. The negative g-factors preclude $m_f = +\frac{1}{2}$ and $+\frac{3}{2}$ as not physically plausible. The g-factor of 0.49 is too small to be likely, but the g-factor of 1.47 is plausibly consistent with gas phase value 1.975. It corresponds to the transition $\left|\frac{1}{2}, -\frac{1}{2}\right\rangle \to \left|\frac{3}{2}, -\frac{1}{2}\right\rangle$, and the absence of other visible excitations implies the selection rule $\Delta m_J = 0$. Consequently, under the assumption that magnetic anisotropy energy of Sm-b is much lower than the Zeeman energy at high field, we consider this transition to be the likeliest assignment of the ~38 mV transition of Sm-b. However, we do not have reason to expect that the anisotropy is smaller than the Zeeman energy, so we next consider cases in which the magnetic anisotropy energy is large for this multiplet of Sm-b.

**Large anisotropy cases**: In the schematic shown in main text Fig. 4e (reproduced in Fig. S13b), we had for simplicity disregarded the zero-field splitting of the $J = 3/2$ excited state that arises from the substrate crystal field. However, this multiplet has non-zero orbital angular momentum ($L = 3$) so significant magneto-crystalline anisotropy is possible. This anisotropy is expected to split the $J = 3/2$ multiplet into an $m_J = \pm\frac{1}{2}$ doublet and a $m_J = \pm\frac{3}{2}$ doublet, as depicted in Fig. S13c,d. If we include large magnetic anisotropy with a leading term describing axial anisotropy $DJ_z^2$, we can no longer use the above simplified expressions (Eqs. S1 and S2) to obtain the $g_{\frac{3}{2}}$ factor. Instead, we consider the Hamiltonian for the Sm electronic spin system:

$$H = g\mu_B \mathbf{J} \cdot \mathbf{B} + D J_z^2 + \lambda \mathbf{L} \cdot \mathbf{S}_{4f} + J_{IA} \mathbf{S}_{6s} \cdot \mathbf{S}_{4f} \qquad \text{(Eq. S3)}$$

where $g\mu_B \mathbf{J} \cdot \mathbf{B}$ is the Zeeman energy of the total moment in the external magnetic field $B$, $DJ_z^2$ is the leading crystal field interaction which gives rise to axial magnetic anisotropy, $\lambda \mathbf{L} \cdot \mathbf{S}_{4f}$ is the spin-orbit coupling between the orbital moment $L$ and $S_{4f}$ that gives rise to the $L$-$S$-tilt excitation energy, and $J_{IA}\mathbf{S}_{6s} \cdot \mathbf{S}_{4f}$ is the intra-atomic exchange interaction between the $6s$ and $4f$ electrons that gives rise to the "$6s$-flip" excitation. The excitation energy is given by the difference $\Delta E = E_{J=3/2} - E_{J=1/2}$, and its shift with magnetic field is $\Delta = \Delta E_{6.5\,T} - \Delta E_0$, which must match our measured shift. The measure a shift is $\Delta = +0.67$ eV, obtained from $\Delta E_0 = 38.36$ mV at $B = 0$ and $\Delta E_{6.5\,T} = 39.03$ mV at $B = 6.5$ T (Fig. 4d).

We make the simplifying assumptions that the spin-orbit term $\lambda \mathbf{L} \cdot \mathbf{S}_{4f}$ and the intra-atomic exchange term $J_{IA}\mathbf{S}_{6s} \cdot \mathbf{S}_{4f}$ are essentially field-independent; in other words, they give the same offsets to $\Delta E_0$ as to $\Delta E_{6.5\,T}$ so they have no effect on the difference $\Delta$, and can be neglected. For the Zeeman energy of $J = \frac{1}{2}$ ground state, we use the directly measured "$m_J$-flip" excitation energy, which for Sm-bv at $B = 6.5$ T is $g_{1/2}\mu_B B/2 = 0.9465$ mV (Fig. 4c). Then we need to compute only the eigenstates of the $J = 3/2$ manifold and use the resulting energy eigenvalues to compute the excitation energies that we match to the experiment. The Hamiltonian for the $J = 3/2$ multiplet is thus simplified to

$$H_{3/2} = g_{3/2}\mu_B \mathbf{J} \cdot \mathbf{B} + DJ_z^2 \qquad \text{(Eq. S4)}$$

Here $g_{3/2}$ is the effective $g$-factor of the $J = 3/2$ multiplet, which arises from the unknown $g$-tensor and our magnetic field direction. This simplified Hamiltonian is intended to capture the main effects of a finite axial magnetic anisotropy term $D$ that tends to orient the spin towards or away from a specific direction. The shift is given by

$$\Delta = E_{3/2}(B = 6.5\,T) - E_{3/2}(B = 0) + g_{1/2}\mu_B B/2 \qquad \text{(Eq. S5)}$$

Here $E_{3/2}$ are the energy eigenvalues obtained from the simplified Hamiltonian (Eq. S4) evaluated at either $B = 0$ T or 6.5 T as shown. We then obtain $g_{3/2}$ and $D$ numerically as the desired fitting parameters. We find that for $D$ oriented along any of the three crystal symmetry directions (normal to the surface plane, along the O-O bridge direction or along the Mg-Mg bridge direction), and with eitherh sign of $D$ with magnitude ranging from 0.20 mV to 10 mV, the result gives a $g$-factor $g_{3/2}$ is less than 1, an implausibly small value, for nearly all cases considered. There are two exceptions, which both require $D < 0$ oriented normal to the surface plane (corresponding to easy-axis anisotropy) (Fig. S13d), so that the external field is oriented mostly orthogonal to the anisotropy axis in our measurement. The first is the situation in which $D$ ranges from $-0.200$ to $-0.275$ mV, for which $g_{3/2}$ ranges from 1.95 to 2.1, corresponding to an excitation $\left|\frac{1}{2}, -\frac{1}{2}\right\rangle \to \left|\frac{3}{2}, -\frac{1}{2}\right\rangle$ with $\Delta m_J = 0$. The second is the situation in which $D$ ranges from $-1$ to $-2$ mV, for which $g_{3/2}$ ranges from 1.6 to 2.0, corresponding to an excitation $\left|\frac{1}{2}, -\frac{1}{2}\right\rangle \to \left|\frac{3}{2}, -\frac{3}{2}\right\rangle$ with $\Delta m_J = -1$. Since only one excitation is observed, both cases imply a selection rule given by the stated $\Delta m_J$, whose justification is beyond the scope of this work to evaluate theoretically, and we consider both cases as plausible explanations of the measurements.

**Consideration of IETS cross sections.** Examination of the IETS cross sections (step heights in d$I$/d$V$) for Sm-b and Eu-b is revealing. For Eu-b, there are two transitions observed in IETS: the $m_J$-changing transition $|4, -4\rangle \to |4, -3\rangle$, which occurs at the Zeeman energy of ~0.74 mV at 6.5 T with conductance change $\Delta\sigma/\sigma \approx 17\%$; and the 6$s$-flip transition $|4, -4\rangle \to |3, -3\rangle$, which occurs at ~163 mV and is much more prominent, with $\Delta\sigma/\sigma \approx 187\%$ (Fig. 1 and Fig. S2). Here $\sigma$ is the elastic conductance, given by the conductance at voltages below the first IETS step. The high efficiency of the 6$s$-flip can be attributed qualitatively to the ease with which the 6$s$ electron alone can be flipped by tunneling electrons. It flips this 6$s$ spin against the entire 4$f$ electron manifold at the expense of an intra-atomic exchange energy penalty[28]. In contrast, the $m_J$-changing transition requires changing $m_J$ for the joint wavefunction of entire 6$s$-4$f$ coupled spin system. Assuming that tunneling electrons interact directly only with the 6$s$ orbital during the inelastic scattering process, as the 4$f$ orbital is shielded from the tunneling electrons, it is reasonable that 6s-flip transitions should be much more efficient than magnetic transitions that involve changes to the f-shell.

Examining the comparable IETS cross sections for Sm-b, the $m_J$-changing transition, which occurs at the Zeeman energy of ~1.8 mV at 6.5 T, has cross section $\Delta\sigma/\sigma \approx 10.5$ %. The 6s-flip transition (at ~148 mV) has IETS cross section $\Delta\sigma/\sigma \approx 140\%$ (Fig. 4). That the IETS cross sections are lower for Sm-b than for Eu-b may be due to the nonzero $L$ of Sm-b, which imparts additional magnetic degrees of freedom for the f-shell electrons involved in the transition that are isolated from the tunneling electrons. The L-S-tilt transition for Sm-b (at ~38 mV) has cross section $\Delta\sigma/\sigma = 56\%$, but there exists no comparable excitation for Eu. Note that the somewhat larger IETS cross section for Eu than for Sm for the 6s-flip excitation was also observed previously for Sm and Eu adsorbed on metal-supported graphene[28]. For each element, we observe a ratio between the 6s-flip and $m_J$-changing transitions of ~11–13, which confirms that these transitions are of similar character.

**Summary.** This section considered alternative interpretations of the L-S-flip excitation at ~38 meV and we conclude that these measurements are consistent with either $\left|\frac{1}{2}, -\frac{1}{2}\right\rangle \rightarrow \left|\frac{3}{2}, -\frac{1}{2}\right\rangle$ and $D = 0$ to $-0.275$ meV, or alternatively $\left|\frac{1}{2}, -\frac{1}{2}\right\rangle \rightarrow \left|\frac{3}{2}, -\frac{3}{2}\right\rangle$ with $D = -1$ to $-2$ meV, in both cases with $D$ oriented normal to the surface plane.

## 6. State assignments for the Sm-O IETS transitions

For Sm on the oxygen binding site, Sm-O, we observed excitations at ~26 mV and ~89 mV (Fig. S12) that we attribute to L-S-tilt transitions, which are excitations into the $J = 3/2$ multiplet ($^8F_{3/2}$). As for Sm-b, we expect that the crystal field will split this multiplet into two doublets, which have $m_J = \pm 1/2$ and $m_J = \pm 3/2$. The diagrams in Fig. S13c–e provide candidate energy orderings. Our spectroscopic data do not include sufficient detail to assign a specific doublet to each excitation energy, but we comment on the possibilities below.

Due to the difficulty of studying this species (for most tip apexes, we were not able to easily reposition Sm to the O-site), we do not have data at elevated bias voltages (>100 mV) as we do for Sm-b. At elevated bias ranges both Sm-b and Sm-O readily hop to the Mg-site, requiring the tip to be backed out increasingly far from the surface as the bias range is elevated to avoid this scenario, which in turn reduces signal/noise. Accordingly in Fig. S12 we obtained tunneling spectra for Sm-O only up to 100 mV, which is not high enough to approach the 6$s$-flip transition energy ~148 mV observed for Sm-b or 188 mV for gas-phase Sm$^+$. Nevertheless, we observe an IETS transition at ~89 mV. Following the above arguments that 6$s$-flip transitions should have the greatest IETS intensity among observable transitions, we assign the ~89 mV excitation, which is weaker than the ~26 mV excitation, to an $L$-$S$-tilt excitation $J = \frac{1}{2} \rightarrow J = \frac{3}{2}$. In this case the higher energy arises from zero-field splitting of the $J = \frac{3}{2}$ multiplet rather than to the 6$s$-flip transition. We point out that in the regime where the magnetic anisotropy energy becomes comparable to the spin-orbit coupling, $J$ and $m_J$ are not good quantum numbers due to state mixing, so that Fig. S13e is oversimplified.

In the discussion for Sm-b above, we could not discern the size of the splitting of the $J = \frac{3}{2}$ states into two doublets, since selection rules appear to prohibit transitions to some of the available excited states. As selection rules can be weakened or modified by the crystal field, and the crystal field environment of Sm-O may differ considerably from that of Sm-b, it is plausible that Sm-O may have different selection rules governing which $J = \frac{1}{2} \rightarrow J = \frac{3}{2}$ transitions are allowed in IETS. We assign the Sm-O spectrum to the excitation scheme shown in Fig. S13e. Due to the C4v symmetry of the oxygen-atop binding site, $D$ is expected to be oriented normal to the surface plane and $E = 0$. Applying a similar analysis using Eq. S3 and S4 as was done for Sm-b, we obtain $|D| = 31$ mV from the splitting of the 26 mV and 89 mV excitations, but cannot determine the sign of $D$ since the expected Zeeman splitting for the $J = 3/2$ state is too small with the external field $B$ oriented nearly orthogonal to $D$. Evidently the entire shift of the ~26 mV excitation at $B = 6$ T seen in Fig. S12 arises from the lowered ground state energy of the $\left|\frac{1}{2}, -\frac{1}{2}\right\rangle$ state, rather than any magnetic field dependence of any of the $J = 3/2$ states.

Alternatively, while the ~26 mV feature is due to a $J = \frac{1}{2} \rightarrow J = \frac{3}{2}$ L-S-tilt excitation, the ~88 mV feature could instead correspond to the $J = \frac{1}{2} \rightarrow J = \frac{5}{2}$ L-S-tilt excitation, which is normally forbidden because $\Delta J = 2$. The $J = \frac{5}{2}$ state is 103.9 mV above the ground state for the gas phase Sm$^+$ ion[35].

Future studies which employ a B field oriented normal to the surface plane (and parallel to D) could conclusively determine the correspondence of the 26 mV and 89 mV excitations to the $J = 3/2$ or $J = 5/2$ states.

## 7. Origin of large Landé g-factors for Sm

This section describes how large Landé g-factors arise, in order to explain the large, measured g-factor (~4.9) for Sm on the bridge site of the MgO film (Sm-b). In most atoms and ions, the Landé g-factor falls between the electron orbital g-factor $g_L = 1$ and the electron spin g-factor $g_e = 2.0023$. Examination of the formula for the Landé g-factor shows that large g-factors arise in atoms and ions for which: (1) S and L are both large, (2) S slightly exceeds L, and (3) the valence shell is less than half full so that Hund's third rule applies to give a value of $J = S - L$ that is small but non-zero. These conditions apply in the case of the ground state of free Sm$^+$ to give $g_J = 4$. Under the approximation that the electron g-factor is $g_e \approx 2$, an atom or ion's Landé g-factor is given by

$$g_J = \frac{3}{2} + \frac{S(S+1) - L(L+1)}{2J(J+1)}$$

For $S > L$, and when J is minimized so that $J = S - L$, we rewrite the g-factor in terms of L and J as

$$g_J = \frac{3}{2} + \frac{(L+J)(L+J+1) - L(L+1)}{2J(J+1)}$$
$$= 2 + \frac{L}{J+1}$$

The plot shown in Fig. S13a illustrates this behavior. This expression shows that minimizing $J$ for a given $L$ maximizes the g-factor. For the smallest positive half-integer value, $J = \frac{1}{2}$,

$$g_J = 2 + \frac{2}{3}L$$

For gas-phase Sm$^+$, we have $L = 3$ and $g_J = 4$, which is the largest known g-factor value for any free atom or ion in its ground state. This limit arises from observing that exceeding $L = 3$ and simultaneously having $S \geq L$ is a condition that would require $S$ to be at least 9/2. This large $S$ might be plausible by having the $f$, $s$, and $d$ subshells all incompletely filled, such as in the hypothetical $4f^6 5d^2 6s^1$ configuration, where the spin is maximized to give $S = 9/2$, and the orbital angular momentum is selected to give $L = 4$. Such a term is not known to exist in any free ion's ground state, but it would theoretically yield the large value $g_J = \frac{14}{3}$.

The g-factors we observe for Sm-b lie in the range 4.71–5.04, which significantly exceeds even the anticipated limit of 4. The analysis above of the Landé g-factor applies when $J$ is a good quantum number, to useful approximation. We expect that the stationary quantum states of atoms in an asymmetric environment (such as adsorbed on surface) will be superpositions of $J$ states. We propose that the large g-factor of Sm-b arises from partial transfer of the 6$s$ electron to the 5$d$ subshell where it can exhibit orbital angular momentum that increases the magnetic moment of the two observed quantum states that are loosely characterized by $m_J = \pm \frac{1}{2}$.

The most extreme negative g-factors may be obtained in cases where $S$ and $L$ nearly cancel but where $L$ exceeds $S$. Minimizing $J$ yields g-factors of largest magnitude. Letting $J = L - S$,

$$g_J = \frac{3}{2} + \frac{(L-J)(L-J+1) - L(L+1)}{2J(J+1)}$$
$$= 2 - \frac{L+1}{J+1}$$

This expression gives a g-factor of $-2/3$, when $L = 3$ as for gas-phase Sm$^+$. In this case $J = 1/2$ and $S = 5/2$ (term $^6F_{1/2}$). We note that the $^6F_{1/2}$ excited state of gas-phase Sm$^+$ (which has a $^8F_{1/2}$ ground state) has a known energy of 188 mV and *g*-factor of −0.595, which is close to this predicted extreme. We observe this excitation at ~148 mV for Sm-b.

## 8. Stabilizing Ln(I) on surfaces

In the main text we proposed that adsorption on metal-supported thin insulating films may serve as a general route to stabilizing open-valence-shell lanthanides in the solid state. Here we discuss this idea further. In previous studies of lanthanides adsorbed on 2 ML MgO grown on Ag(001), individual atoms such as Ho, Dy and Er underwent charge transfer to the underlying substrate to form a +1 cation[3,24,28], as did Eu and Sm in the present work. A key property in determining whether a given element shows an open valence shell cation is its propensity to form a divalent configuration in compounds, as in Eu and Sm, versus a trivalent configuration, as in Ho, Dy and Er, which instead promote a 4*f* electron to the 5*d* shell. The combined result of this 4*f*-5*d* electron promotion and cation formation is a closed-shell $6s^2$ valence configuration, with the 5*d* electron having transferred to the underlying substrate. For Eu and Sm, the stronger preference for a divalent configuration, and the consequent lack of 4*f*-5*d* electron promotion, means that cation formation instead results in the desired $6s^1$ open-shell valence configuration, which we propose is the key to electronic access to their magnetic properties.

We note here several aspects that stimulate further investigation:

1) We expect that other lanthanides with a preference for a divalent electron configuration, such as Yb and Tm[30], should also form monovalent cations on 2 ML MgO/Ag(001) similar to Eu and Sm.

2) As noted in refs.[51–53], ultra-thin insulating films (< 5 ML) modify the work function of the surface, facilitating charge transfer to or from adsorbates. Different combinations of insulating film material, thickness, and underlying conductor influence the propensity of adsorbates to undergo charge transfer, so they offer a rich state space to explore for studies of open-shell Ln(I).

3) Most lanthanides studied in the graphene-on-metal adsorption environment showed easily detected magnetic excitations in IETS, so they were also likely monovalent[28]. In experiments on borozene clusters, lanthanides having a divalent preference such as Tm and Yb, in addition to those with stronger trivalent preference (La, Pr, but notably not Tb) could similarly be stabilized in $6s^1$ monovalent configurations[27]. These observations suggest that details of the local adsorption environment influence whether individual Ln atoms promote $4f$ electrons to the $5d$ shell, in addition to their element-specific propensities. We can expect that other thin insulating films aside from MgO may result in a greater or lesser propensity for adsorbed lanthanides to undergo $4f$-$5d$ promotion. These other thin films may allow elements other than Eu, Yb, Sm and Tm to be stabilized in the $6s^1$ monovalent configuration, including some of the more trivalent-preferring lanthanides.

The above considerations warrant further theoretical and experimental investigation in understanding what factors influence $4f$-$5d$ electron promotion in a surface adsorption environment, which is distinct from most other chemical environments due to the under-coordination of the Ln ion. As we have seen in the present work, there are also variations in charge or valence configuration which depend on the particular adsorption site on a given substrate, exemplified by the change to a closed-shell configuration when Eu and Sm are adsorbed on the Mg-site. We anticipate that a better developed understanding of these factors will lead to more studies that take advantage of the unique chemical and magnetic properties of monovalent lanthanides.

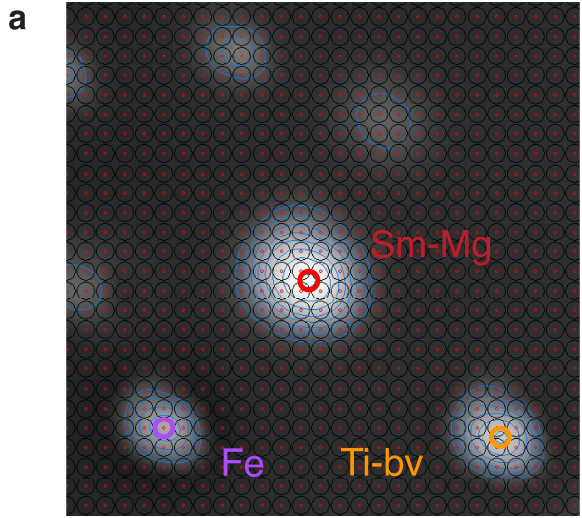
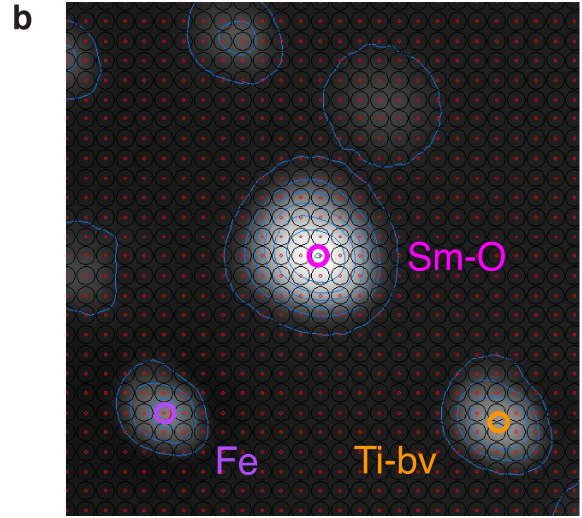
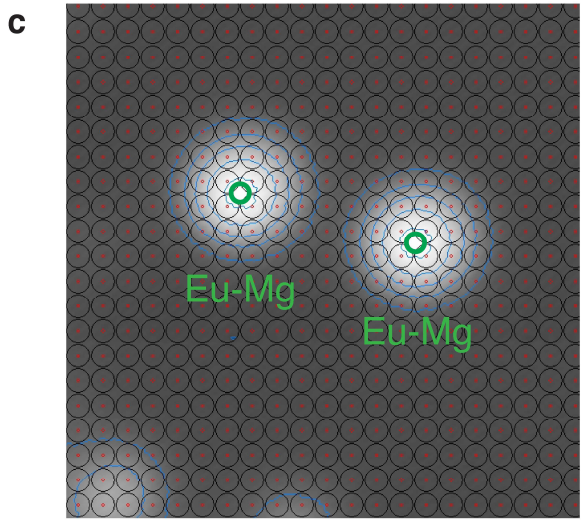
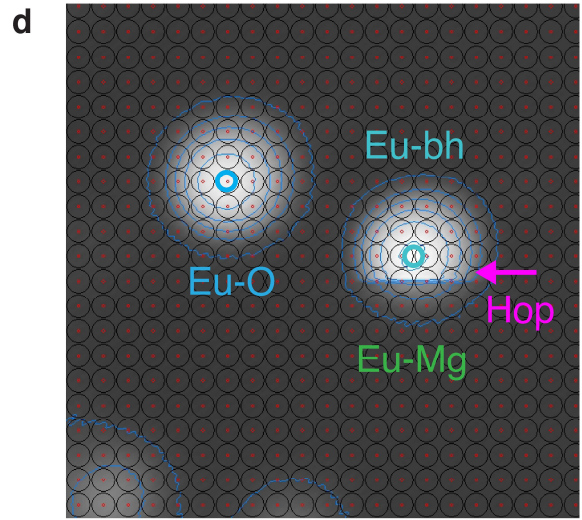
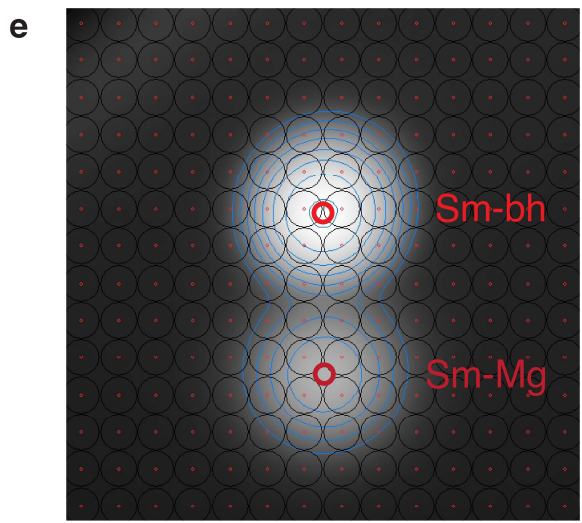
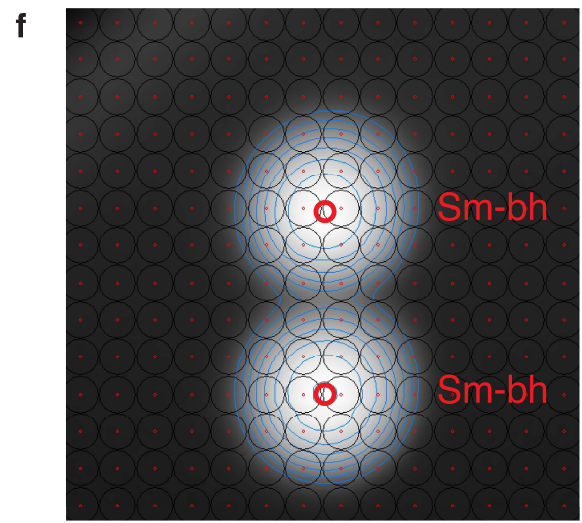

**Fig. S1. STM images and binding sites for Eu and Sm on MgO film.**

**a**, Binding site of Mg-site Sm (Sm-Mg) depicted with underlying MgO lattice (red points represent oxygen atoms). Nearby oxygen-site Fe (Fe-O) and vertical-bridge-site Ti (Ti-bv) atoms are used to align the lattice. **b**, Binding site of Sm-O after using atom manipulation to reposition the Sm atom in **a**. **a** and **b** were acquired with current setpoint $I_{set}$ = 15 pA at $V_{set}$ = 50 mV, image size 7.5 nm square. **c**, Binding site of Eu-Mg atoms with lattice overlay. **d**, Binding sites of Eu-O and horizontal-bridge-site Eu (Eu-bh) after using atom manipulation to reposition the rightmost Eu-Mg to the bh site. Unusually, this tip apex would concurrently reposition the neighboring Eu-Mg atom (left) to the O-site when performing atom manipulation with the tip apex centered on the rightmost Eu. The rightmost Eu atom can be seen to hop at the labeled scan line from the bh site to the Mg site during image acquisition. **c** and **d** acquired with 5 pA at 50 mV, image size 6 nm square. **e,f**, Binding sites of Sm-bh and Sm-Mg before (**e**) and after (**f**) atom manipulation used to reposition the bottom Sm atom to the bh site. **e** and **f** acquired with 10 pA at 50 mV, image size 4 nm square.

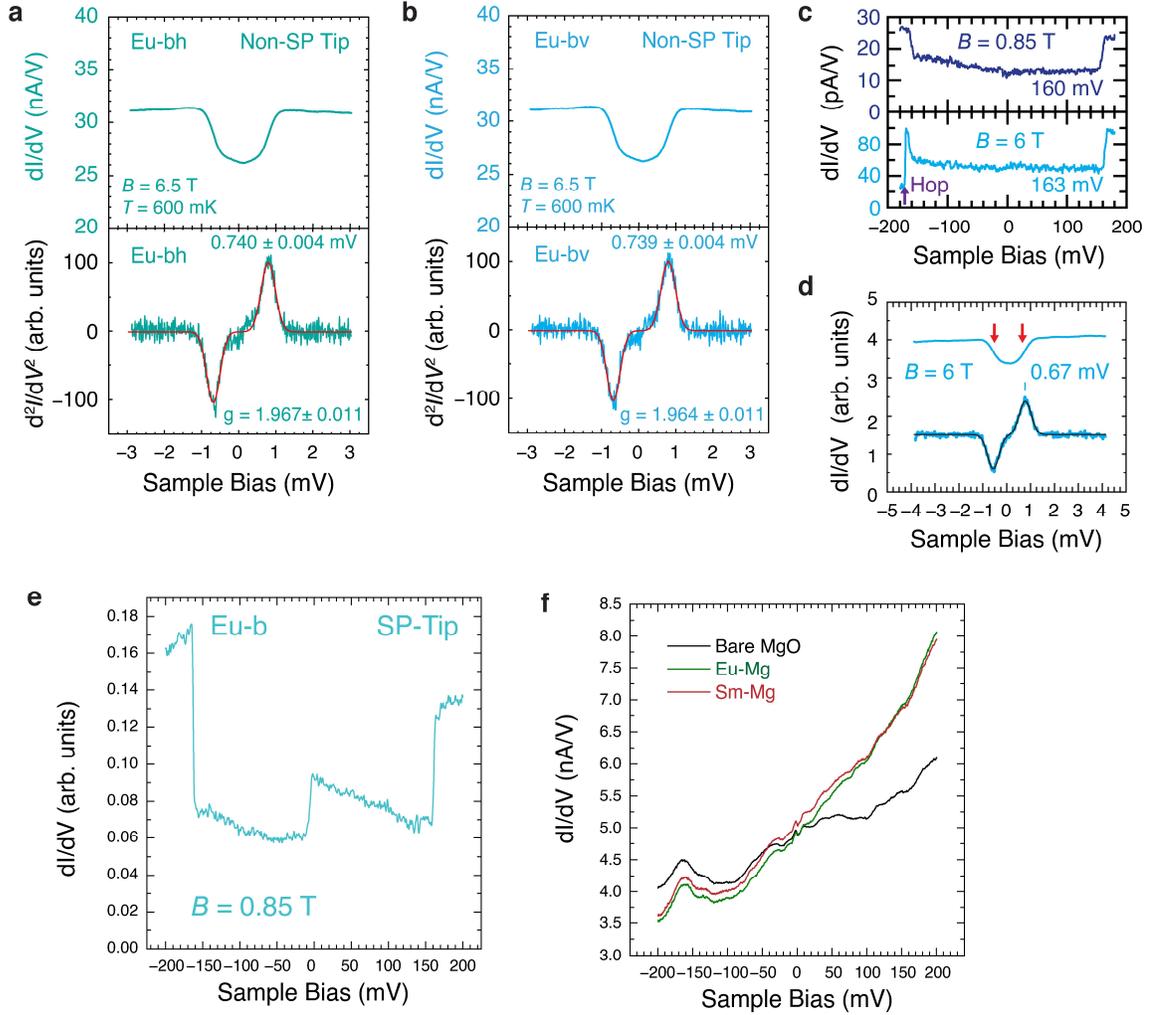

**Fig. S2. Additional tunneling spectra of bridge-site Eu (Eu-b).**

**a**, Tunneling spectra of horizontal bridge-site Eu (Eu-bh), acquired at tip-height setpoint 300 pA at 10 mV, AC modulation $V_{AC} = 100$ µV, $B = 6.5$ T, $T = 600$ mK. **b**, Tunneling spectra of vertical bridge-site Eu (Eu-bv), same conditions as **a**, reproduced from main text Fig. 1f. In **a** and **b**, a Gaussian fit to each peak, averaged over positive and negative bias, gives energy as labeled. No difference in the excitation energy is discerned between the two orientations, bh and bv, providing evidence that Eu-b has negligible in-plane magnetic anisotropy and high in-plane symmetry of the $g$-tensor. **c**, Tunneling spectra of Eu-b acquired using two additional tip apexes at 0.85 T (top) and 6 T (bottom). Each spectrum was acquired with a unique tip apex which differed from the tip apex used in Fig. 1e of the main text. Top: tip-height setpoint $I_{set} = 2.5$ pA at $V_{set} = 80$ mV. Bottom: setpoint 5 pA at 80 mV. For both spectra, $V_{AC} = 3$ mV rms, $T = 600$ mK. **d**, Low bias tunneling spectra of Eu-b showing the Zeeman splitting at 6 T, acquired with feedback setpoint 200 pA at 10 mV, $V_{AC} = 200$ µV, $T = 600$ mK. Top trace: $dI/dV$; bottom trace: $d^2I/dV^2$ numerically computed. **e**, Tunneling spectra of Eu-b acquired with a spin-polarized tip, setpoint 40 pA at 20 mV. Note the abrupt conductance step near zero bias and the large conductance step height asymmetry between positive and negative bias for the 163 mV excitation. These asymmetries arise from spin-polarized IETS selection rules and further evidence the magnetic nature of both the zero-bias excitation and the 163 mV excitation of Eu-b[34]. **f**, Tunneling spectra of Eu-Mg, Sm-Mg and the bare MgO substrate as a background measurement. No IETS features were observed for either Eu-Mg or Sm-Mg, in contrast to b-site and O-site adsorbed species. All spectra in **f** were acquired with tip-height setpoint 50 pA at 10 mV, $V_{AC}$ = 1 mV, $B$ = 0 T.

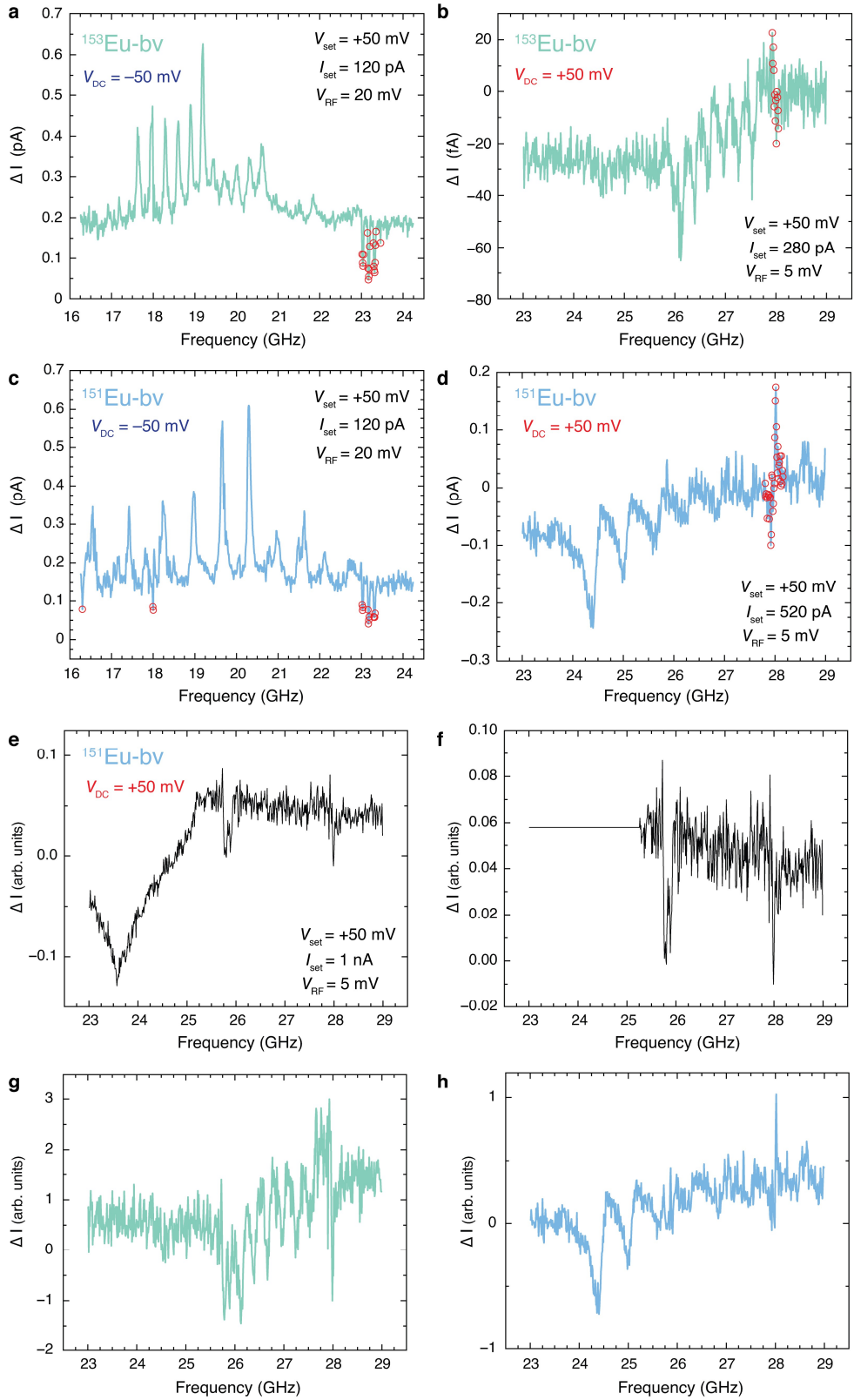

**Fig. S3. Raw data for ESR spectra of Eu-b shown in main text Fig. 2.**
**a,b**, ESR spectra for $^{153}$Eu. **c,d**, ESR spectra for $^{151}$Eu. For **a**–**d**, the spectra are the same data as those shown in the main text Fig. 2, here plotted as connected lines instead of points and without peak fits. Here the entire spectrum is shown including points dominated by artifacts that arise from our instrument's uncompensated frequency-dependent transfer function at certain frequencies (marked in red). The red points were omitted in the main text for clarity. **e**, ESR spectrum for $^{153}$Eu acquired at tip-height setpoint $I_{set}$ = 1 nA at $I_{set}$ = 50 mV, $V_{RF}$ = 5 mV 0-p (zero-to-peak). The closer tip proximity redshifted the ESR features toward 24 GHz, so that the 25.5–29.5 GHz frequency region was free of ESR signals so it could be used as a background measurement to subtract artifacts due to the uncompensated transfer function. **f**, Same as **e** but with the frequency range containing ESR signals (< 25.5 GHz) removed and replaced with a constant background. The spectrum in **f** was subtracted from the raw data to obtain the main text Fig. 2 spectra acquired at positive bias, and to obtain spectra in **b** and **d** shown here. **g,h**, Raw data for $^{153}$Eu and $^{151}$Eu, respectively, before any background subtraction.

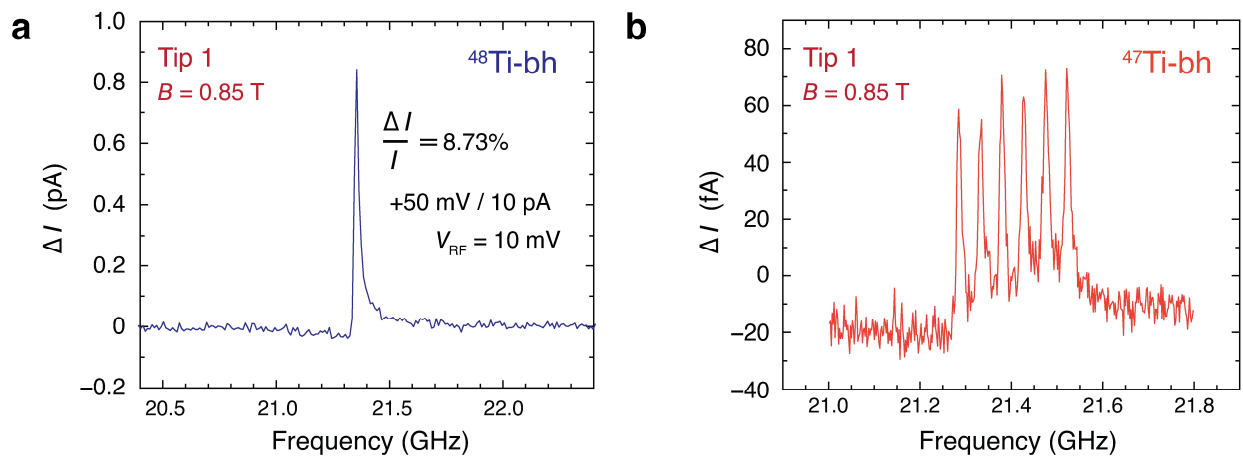

**Fig. S4. ESR spectra of Ti isotopes $^{48}$Ti-b and $^{49}$Ti-b.**
**a**, ESR spectrum of $^{48}$Ti-b (nuclear spin $I = 0$) at $B = 0.85$ T. The strong signal/noise of the apex used to acquire this data resulted in an ESR peak amplitude of over 800 fA, which represents a change in tunneling current of ~8.7 % on resonance. **b**, ESR spectrum of $^{47}$Ti-b ($I = 5/2$, giving 6 peaks) acquired with the same tip apex under similar conditions, averaged over 4 passes. The peak heights are less than one sixth of those for the zero nuclear spin isotope because each peak corresponds to one of the six thermally occupied nuclear spin states. **a** and **b** use "Tip 1", the same tip apex used to acquire all ESR data shown in the main text Figs. 2, 3 and 5, and supplementary Figs. S3, S4, S6, S7, S14d–g, and S15. Since the Eu ESR signal intensity of each peak is significantly reduced owing to the large number of electron and nuclear spin states, the quality and signal-to-noise ratio for magnetic tips needs to be especially high for studying Eu. We provide the Ti-b spectra here as a reference for this tip's performance.

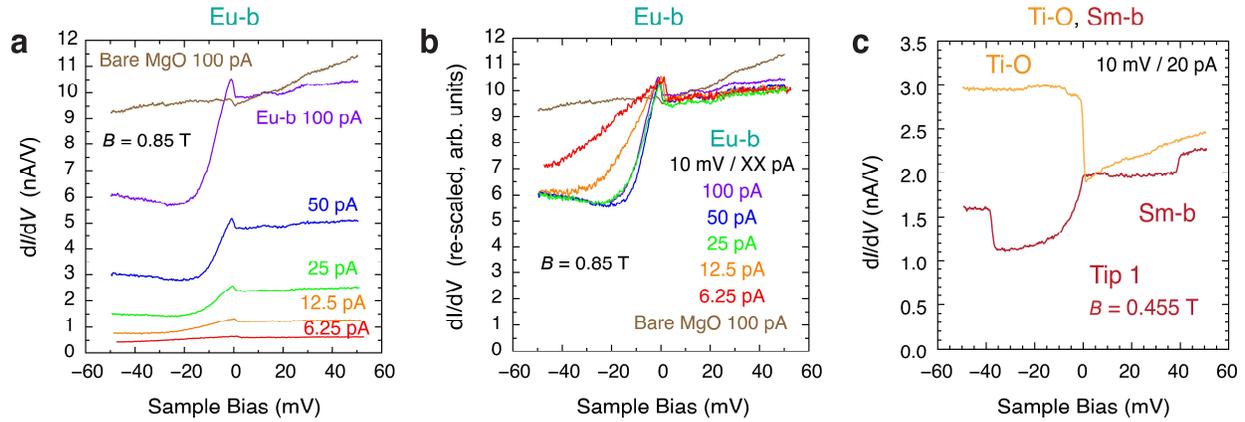

**Fig. S5. Spin torque of Eu-b and detection of tip spin polarization using Ti-O.**
**a**, Tunneling spectra (d$I$/d$V$) of Eu-b at varying tip-surface distance, established by varying the tip-height setpoint current as indicated with each trace, at $V_{set}$ = 10 mV, $V_{mod}$ = 0.5 mV rms, $B$ = 0.85 T. All traces shown were acquired with the tip positioned over Eu-b except the brown trace which was acquired with the tip positioned over the bare MgO film as a background measurement at a position laterally far from any adsorbate. **b**, Data shown in **a** with each trace re-scaled vertically by normalizing to the $I$ = 100 pA trace based on the ratio of setpoint currents. Negative-bias dropoff of conductance as current is increased indicates spin torque. **c**, Tunneling spectra of Sm-b and Ti-O acquired with the same tip apex at setpoint 20 pA at 10 mV, $B$ = 0.455 T, $T$ = 0.6 K. The conductance change near zero bias for Ti-O arises from spin-polarized inelastic excitation selection rules and can be used to characterize the spin polarization of the tip. The tip used here, "Tip 1", was the same tip apex used to acquire all ESR data in main text: Figs. 2, 3 and 5, and supplementary Figs. S3, S4, S6, S7, S14d–g and S15.

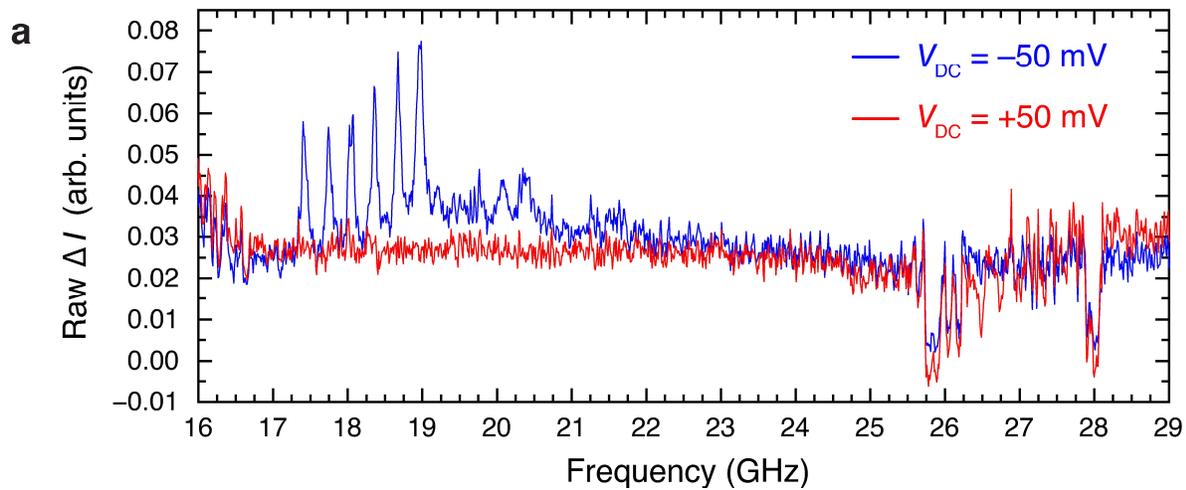

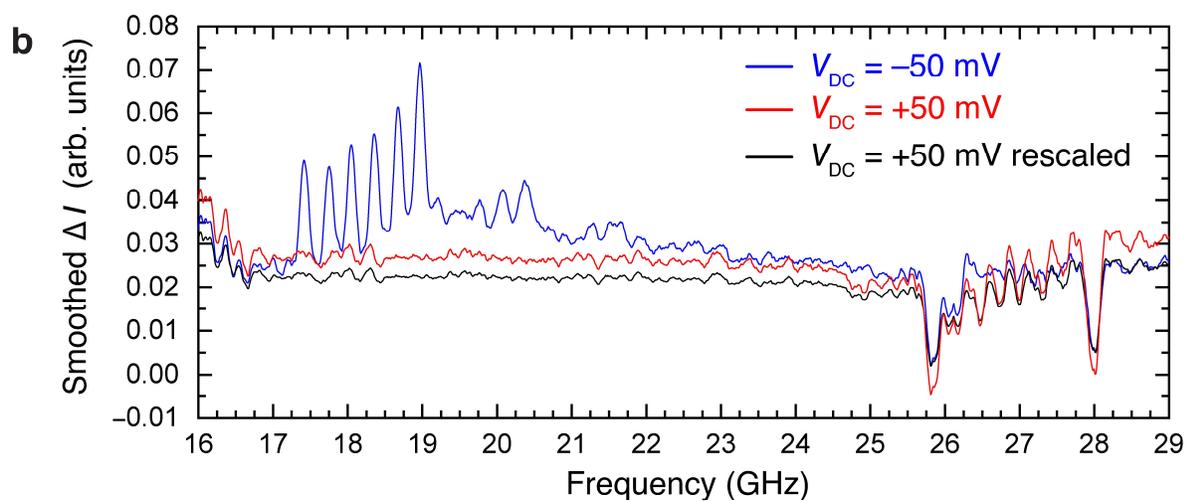

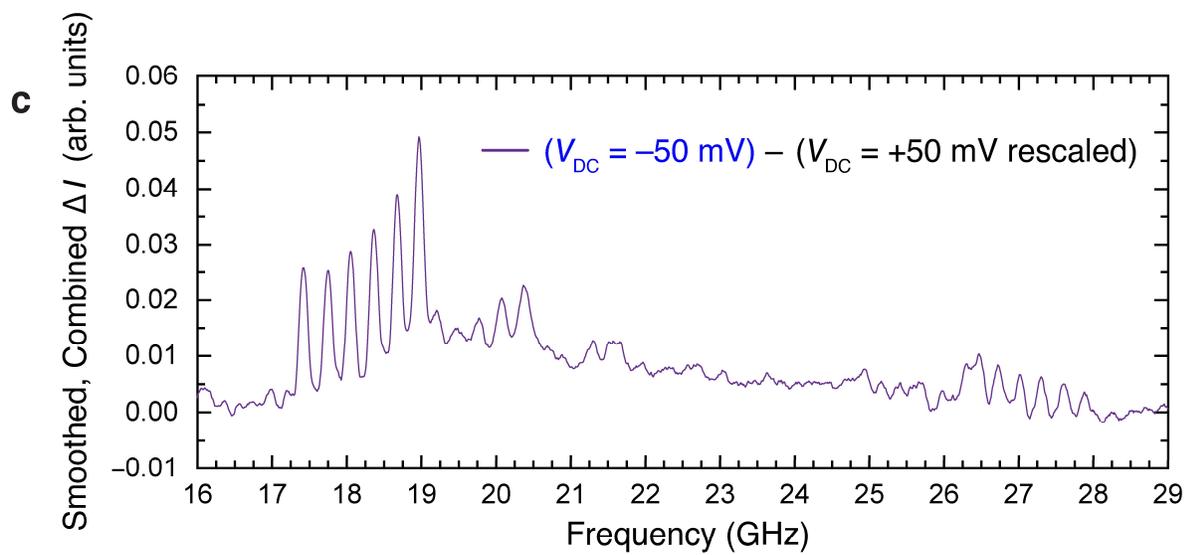

**Fig. S6. Processing ESR spectra for $^{153}$Eu-b shown in main text Fig. 3.**

**a**, ESR spectra acquired with the tip positioned over a $^{153}$Eu-bv atom (bridge-site in vertical direction) with tip-height setpoint $I_{set}$ = 67 pA at $V_{set}$ = −50 mV (blue curve) and $I_{set}$ = 200 pA at $V_{set}$ = +50 mV (red curve); $B$ = 0.85 T, $V_{RF}$ = 10 mV. These two setpoints are approximately identical in tip height (tip-surface distance) despite the difference in tunnel junction conductance because of the larger magneto-conductance at positive bias found for this junction. **b**, Positive and negative bias spectra from **a** smoothed using the Savitzky-Golay method with a 20-point averaging window. The black curve corresponds to re-scaling the red (positive bias) curve by a factor 0.65 to approximately match the signal strength of the rectification background and transfer-function anomalies present in the negative bias (blue) curve. **c**, Subtracted, smoothed data, also shown as Fig. 3a of the main text. The rescaled, smoothed positive bias data (black curve in **b**) was subtracted from the smoothed negative bias data (blue curve in **b**) to produce the complete spectrum shown in purple. Since the ESR signal at positive and negative bias had opposite amplitude, subtracting the positive bias data (where ESR signals are dips) from the negative bias data (where ESR signals are peaks) produces a spectrum that shows ESR features from both bias polarities as peaks while largely cancelling the transfer function artifacts present in both original spectra. This method allows a more complete view of the $^{153}$Eu ESR spectrum, by combining the data resulting from spin-torquing the electron spin toward the ground state (positive bias) and spin-torquing toward the most excited states (negative bias).

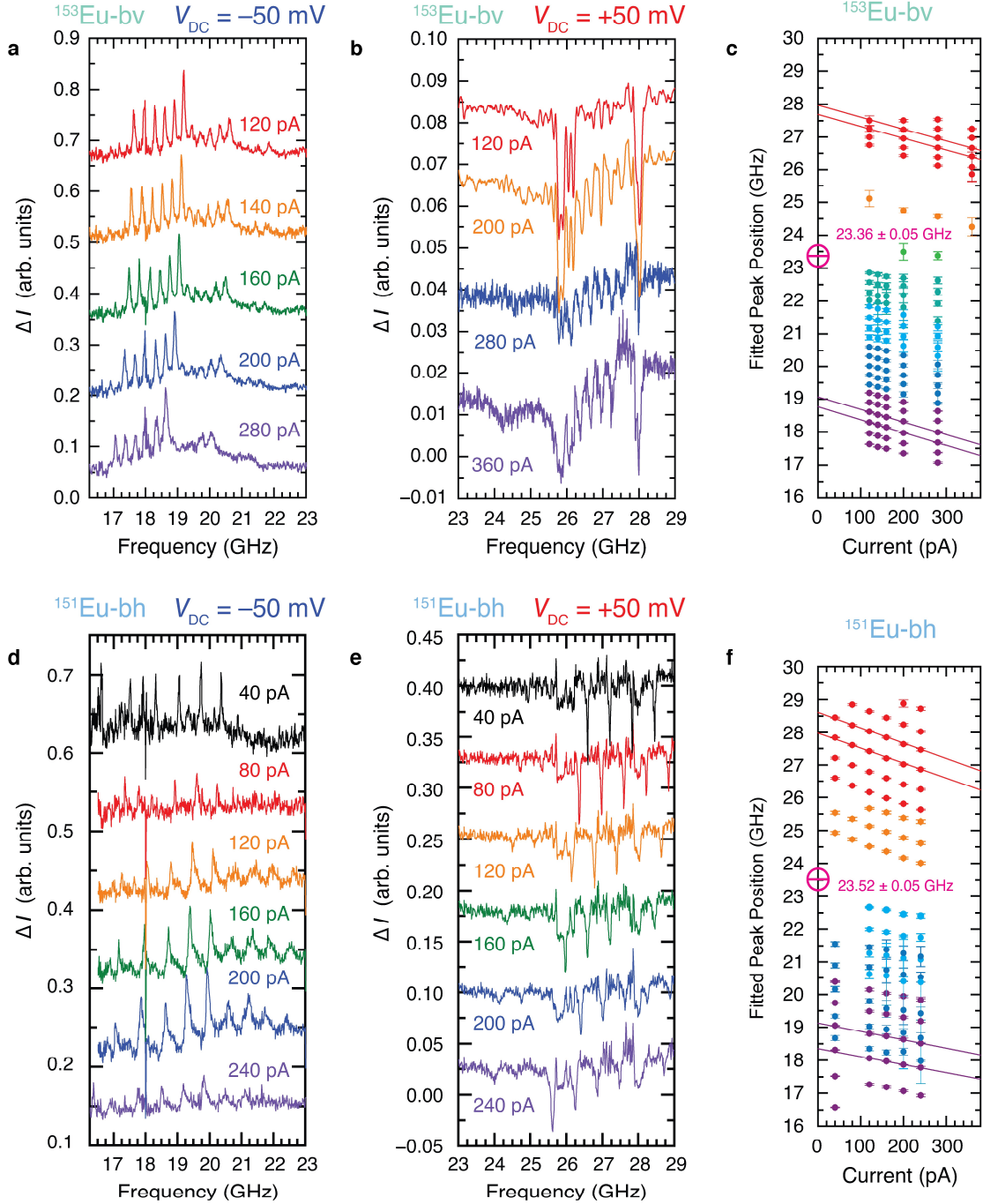

**Fig. S7. Determining the *g*-factor of Eu-b from ESR spectra.**
**a**, ESR spectra for $^{153}$Eu-bv acquired at different tip heights, established by varying the setpoint tunneling current as labeled at $V_{set}$ = +50 mV. Bias voltage during the ESR acquisition was $V_{DC}$ = −50 mV. **b**, ESR spectra acquired at same conditions except $V_{DC}$ = +50 mV. **c**, Fitted peak positions for all spectra shown in **a** and **b**. Linear fits to the peak positions for tree 1 peaks 3 and 4 (red lines) and tree 8 peaks 3 and 4 (purple lines) are shown. The y-intercepts of these four linear fits were then averaged to approximate the Zeeman energy (pink circle on vertical axis) of the atom with the effects of the crystal field and tip magnetic field mostly cancelled. **d**–**f**, Similar ESR spectra and resulting fitted peak positions for a different isotope and bridge orientation, $^{151}$Eu-bh. For all panels, $B = 0.85$ T and $V_{RF} = 20$ mV for negative bias spectra, $V_{RF} = 10$ mV for positive bias spectra. All spectra were acquired with feedback loop open after setting the tip height using the indicated setpoint currents at $V_{set} = +50$ mV. For negative bias spectra the feedback was established at positive bias and then the bias was changed to $V_{DC} = -50$ mV under open loop conditions before data acquisition. (An exception was the 40 pA spectrum for $V < 0$ in which the bias was instead ramped to $V_{DC} = -100$ mV to improve signal strength; this data point was not included in the linear fits). The extrapolated Zeeman energy provides an estimate of the electron $g$-factor, giving $1.964 \pm 0.004$ for $^{153}$Eu-bv and $1.977 \pm 0.004$ for $^{151}$Eu-bh. These error bars arise from the errors in the peak fits, but do not take into account systematic uncertainties in the total magnetic field which includes a tip dipole field. This tip dipole field is uncharacterized and can be aligned along or orthogonal to the in-plane transverse anisotropy $E$ direction, which may account for the difference in apparent $g$-factors between the two measurements shown. See Supplementary Note 2 for further discussion. Spectral features near 18 GHz, 25.75 GHz and 28 GHz are artifacts due to the microscope's uncompensated frequency-dependent RF transfer function.

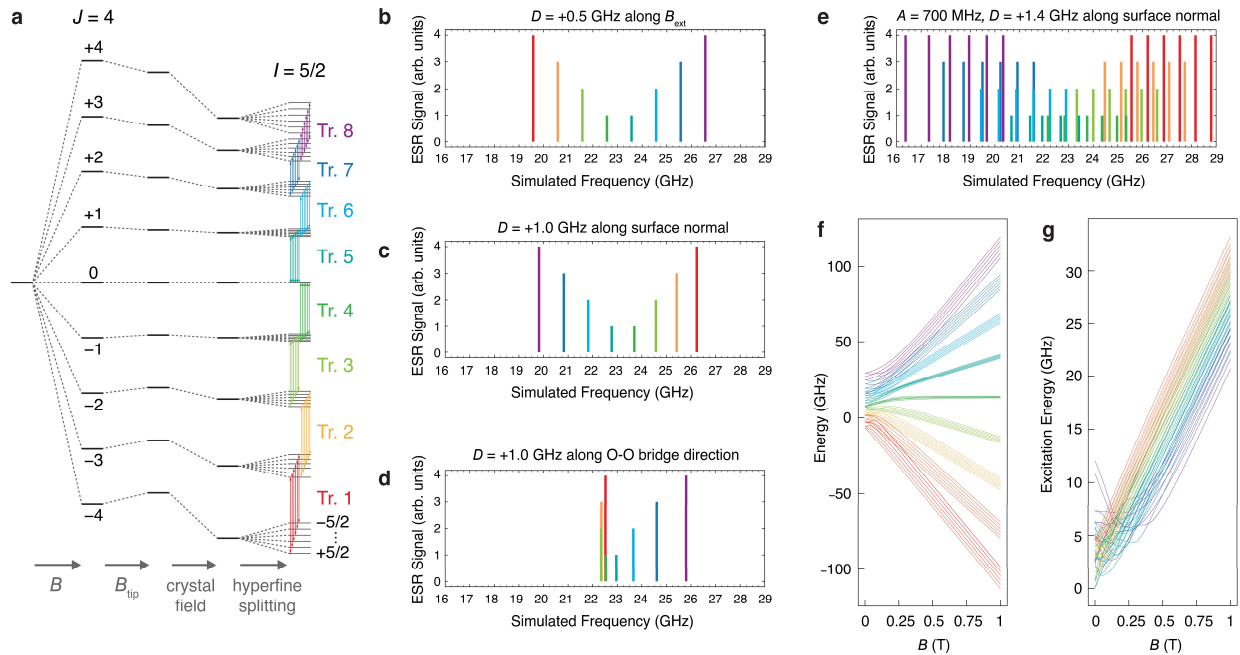

**Fig. S8. Calculated states and transitions from model Hamiltonian of Eu-b.**
**a**, Schematic energy level diagram showing the qualitative effect of each term of the model Hamiltonian, for the $J = 4$, $I = 5/2$ spin system of Eu$^+$. The effects of each term are not drawn to scale. **b**, Simulated excitation energies assuming an axial magnetic anisotropy term $D = +0.5$ GHz oriented along the external magnetic field with $B = 0.85$ T. For illustration, hyperfine interactions are excluded here, so that there are 9 magnetic states (8 electron spin-flip transitions). The depicted height of each excitation is arbitrary and chosen to be suggestive of heights seen in measured spectra. Colors depict the ordering of the initial energy eigenstate, with red corresponding to excitations between the ground state and first excited state, and purple corresponding to excitations between the second-most and most excited energy state. **c**, Simulated excitation energies for $D = +1.0$ GHz oriented normal to the surface plane, assuming $B = 0.85$ T oriented mostly in-plane with a 9° tilt toward the surface normal. Note the reversed excitation energy ordering relative to **a** that results. The energy spacing between excitations is no longer equal, with a slightly larger spacing occurring between the higher energy states (which are lower-energy transitions, shown in purple, blue etc.) compared to states near the ground state (red, orange etc.) This anisotropy axis and parameter gives qualitative agreement with measured Sm-b spectra. **d**, Simulated excitation energies for $D = +1.0$ GHz oriented the O-O bridge direction, assuming an in-plane magnetic field oriented 45° relative to the O-O direction. The first four excitation energies are seen to be nearly degenerate and thus not consistent with the measured Sm-b spectra. **e**, Simulated ESR spectrum for hyperfine interaction strength $A = +700$ MHz, $D = +1.4$ GHz oriented along the surface normal similar to **c**, with $B = 0.84$ T. These parameters are intended to qualitatively model the measured spectrum of $^{151}$Eu shown in Fig. S7d,e. Colors depict excitations between electron spin states, with all transitions for a given nuclear spin state colored the same. **f**, Energy eigenvalues and **g**, excitation energies for the model parameters used in **e** as a function of $B$. As $B$ approaches zero, the spin states are no longer well-described as a product of electron and nuclear spin states, and instead the total angular momentum $F = I + J$ becomes a good quantum number for stationary states.

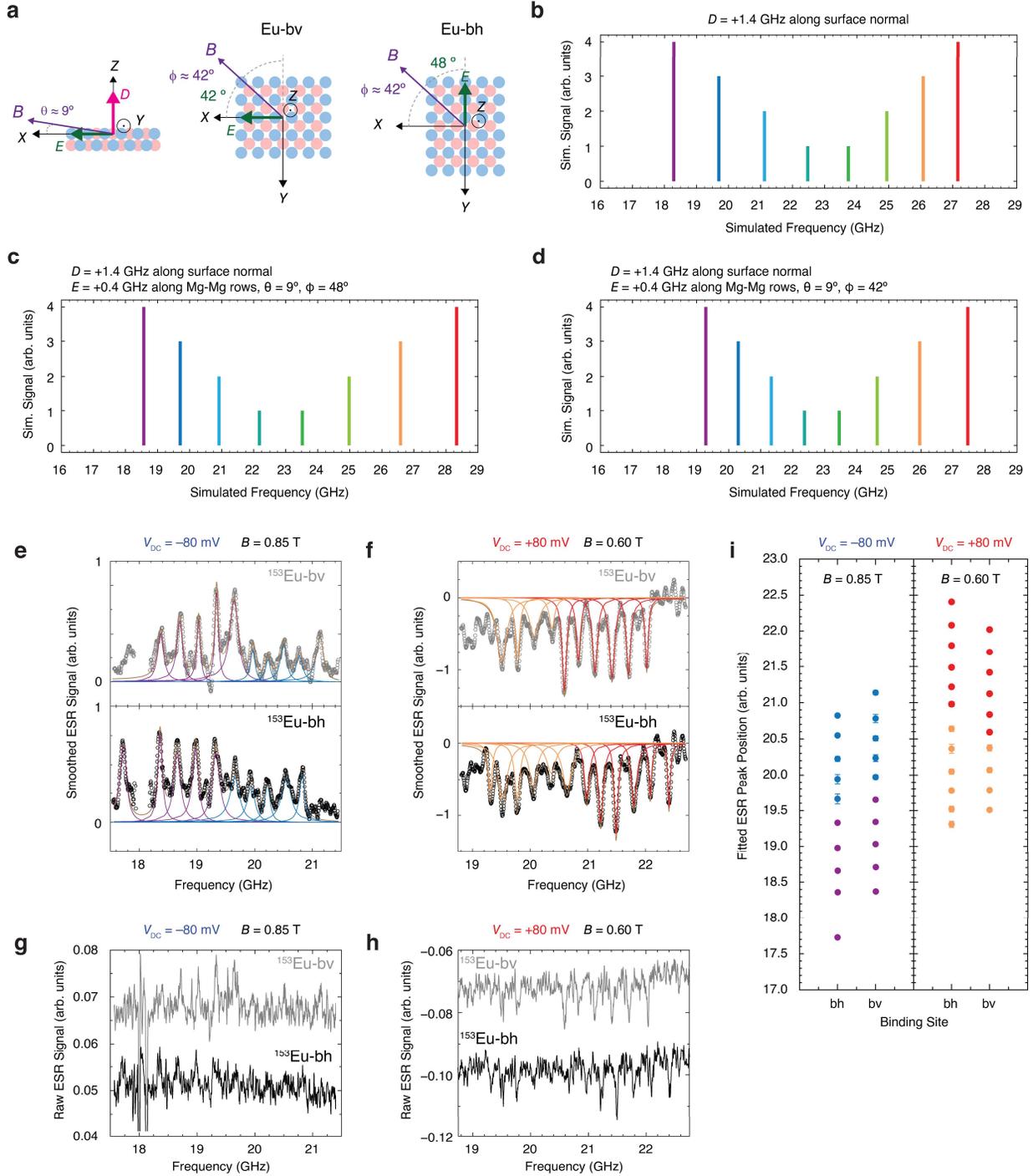

**Fig. S9. Effect of transverse crystal field *E* on model Hamiltonian for Eu-b.**

**a**, Schematic diagram depicting the orientation of the external magnetic field and Eu-b crystal field axes (*X, Y, Z*) with respect to the crystal surface. Left: side view showing the external field *B*, its out-of-plane angle θ and the axial anisotropy term *D*. Center and right: top-down views of Eu-bv and Eu-bh respectively, with the external magnetic field vector (purple arrow), in-plane angle φ, and transverse magnetic anisotropy term *E* (green arrow), which is oriented along the Mg-Mg row direction. **b**, Model Hamiltonian simulation of the excitation spectrum for Eu assuming $D = +1.4$ GHz oriented along the crystal $Z$ axis, $\theta = 9°$ and $E = 0$, with the hyperfine coupling constant $A$ is set to zero for simplicity. **c**, Same as **b** except $E = +0.4$ GHz oriented along the Mg-Mg direction and $\varphi = 48°$. **d**, Same as **c** except $\varphi = 42°$. These values for $\varphi$ correspond approximately to the experimental external magnetic field orientations with respect to $E$ for the two bridge directions. **e**, Smoothed ESR spectrum of Eu-bv (top, gray) and Eu-bh (bottom, black) acquired at $V_{DC} = -80$ mV, tip-height setpoint $I_{set} = 40$ pA and $V_{set} = +80$ mV; $V_{RF} = 30$ mV, $B = 0.85$ T. **f**, Smoothed ESR spectrum of Eu-bv and Eu-bh acquired at $V_{DC} = -80$ mV, $I_{set} = 40$ pA, $V_{set} = +80$ mV, $V_{RF} = 30$ mV, $B = 0.60$ T. Here $B$ was chosen to reduce the frequeency of peaks in trees 1 and 2 to the most transmitting RF window of our instrument transfer function, so that they appear in a similar range as trees 7 and 8 in **e**. **g,h**, Raw ESR spectra of **e** and **f**, without smoothing, peak fits, or removal of artifact-dominated points. **i**, Fitted peak positions for the data shown in **e** and **f**. The redshift of excitations for Eu-bh in trees 7 and 8 (purple and blue points) and blueshift of excitations for Eu-bh in trees 1 and 2 (red and orange points) relative to comparable peaks for Eu-bv is predicted by the transverse magnetic anisotropy term $E$ as shown in **c**, and **d**. In contrast, anisotropy in the Landé $g$-factor along the Mg-Mg and O-O bridge directions would cause the peaks of all trees for Eu-bh to redshift or blueshift compared to Eu-bv. In other words, the presence of transverse anisotropy $E$ causes the two different Eu-b adsorption sites to appear to have different crystal field splitting when $\varphi$ differs significantly from 45°. In contrast, $g$-factor anisotropy would cause the two different Eu-b adsorption sites to have the same crystal field splitting, and an overall frequency offset for all peaks caused by the different effective $g$-factors. The presence of an out-of-plane axial anisotropy $D$ without a transverse anisotropy $E$ would cause Eu-bv and Eu-bh to have identical spectra.

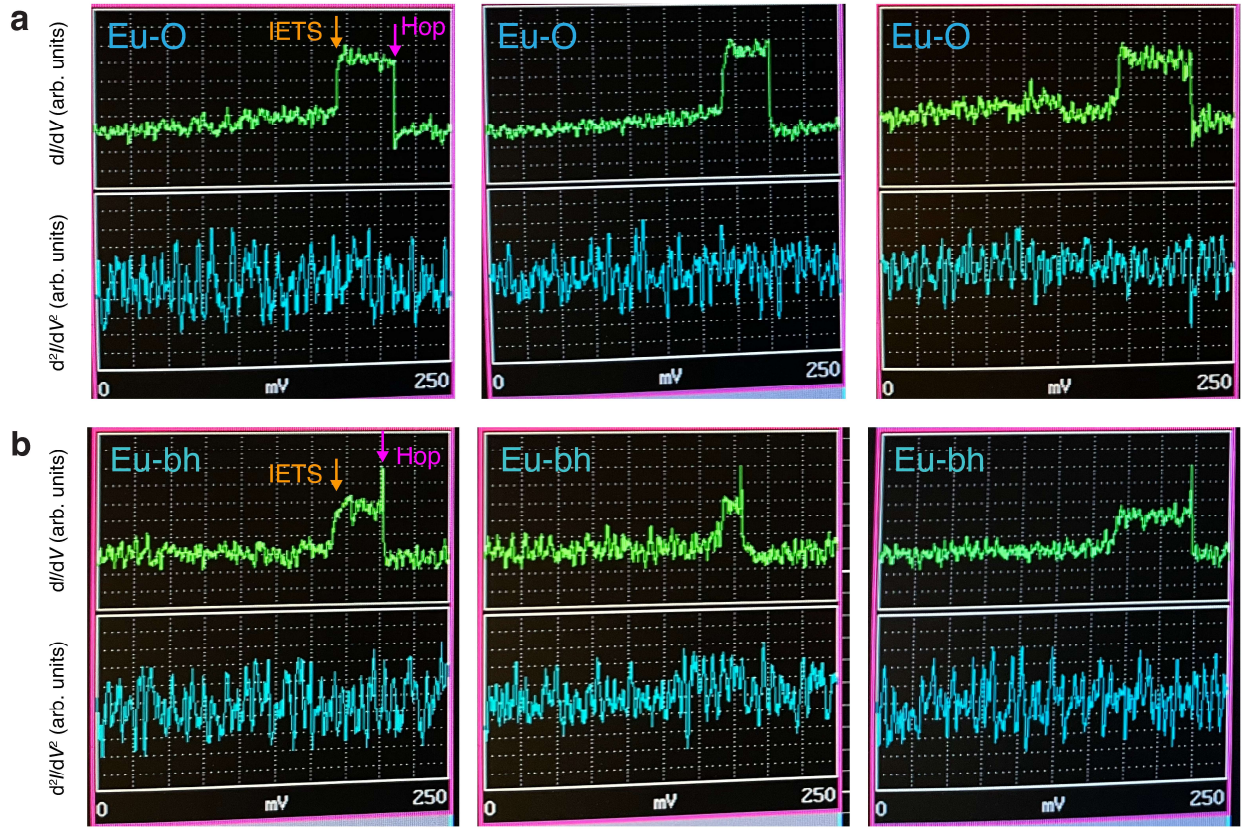

**Fig. S10. Tunneling spectra of Eu at oxygen and bridge sites.**

**a,b**, Three d$I$/d$V$ traces each for Eu-O (**a**, top row) and Eu-bh (**b**, bottom row). Top (green) traces in each panel are d$I$/d$V$. Bottom (blue) traces are d$^2I$/d$V^2$, which are not reliably distinguishable from noise in these rapidly acquired spectra. The upward conductance step in d$I$/d$V$ (~163 mV for Eu-b and ~165 mV for Eu-O) is the "6$s$-flip": the intra-atomic spin-flip excitation shown in main text Fig. 1e,g. The downward conductance steps at ~180–220 mV result from the Eu atom hopping to the lower conductance Mg site because of the elevated bias voltage. The d$I$/d$V$ traces shown here were acquired using a fast d$I$/d$V$ acquisition mode of our instrument that enables rapid characterization of atoms, but which does not store the resulting data. Here we show photographs from the data-acquisition computer display. These spectra lead us to assign an open-shell (6$s^1$) configuration to Eu-O and to Eu-b, where the conductance steps visible here correspond to 6$s$-flip excitations, as for Sm-O and Sm-b. For Eu-O and Eu-b this transition reduces total spin $S$ from 4 to 3 (term $^9S_4$ to $^7S_3$), as for the free Eu$^+$ ion excitation at 207 mV. Regrettably, this tip apex was lost before quantitative data could be recorded for Eu-O, and this particular tip was the only tip apex with which we could readily move Eu to the O-site. We show this limited data to provide qualitative evidence that Eu-O is in the same open valence shell configuration as Eu-b, similar to Sm-O and Sm-b respectively. The Eu-O atom studied here was the same as that shown in the topograph in Fig. S1d.

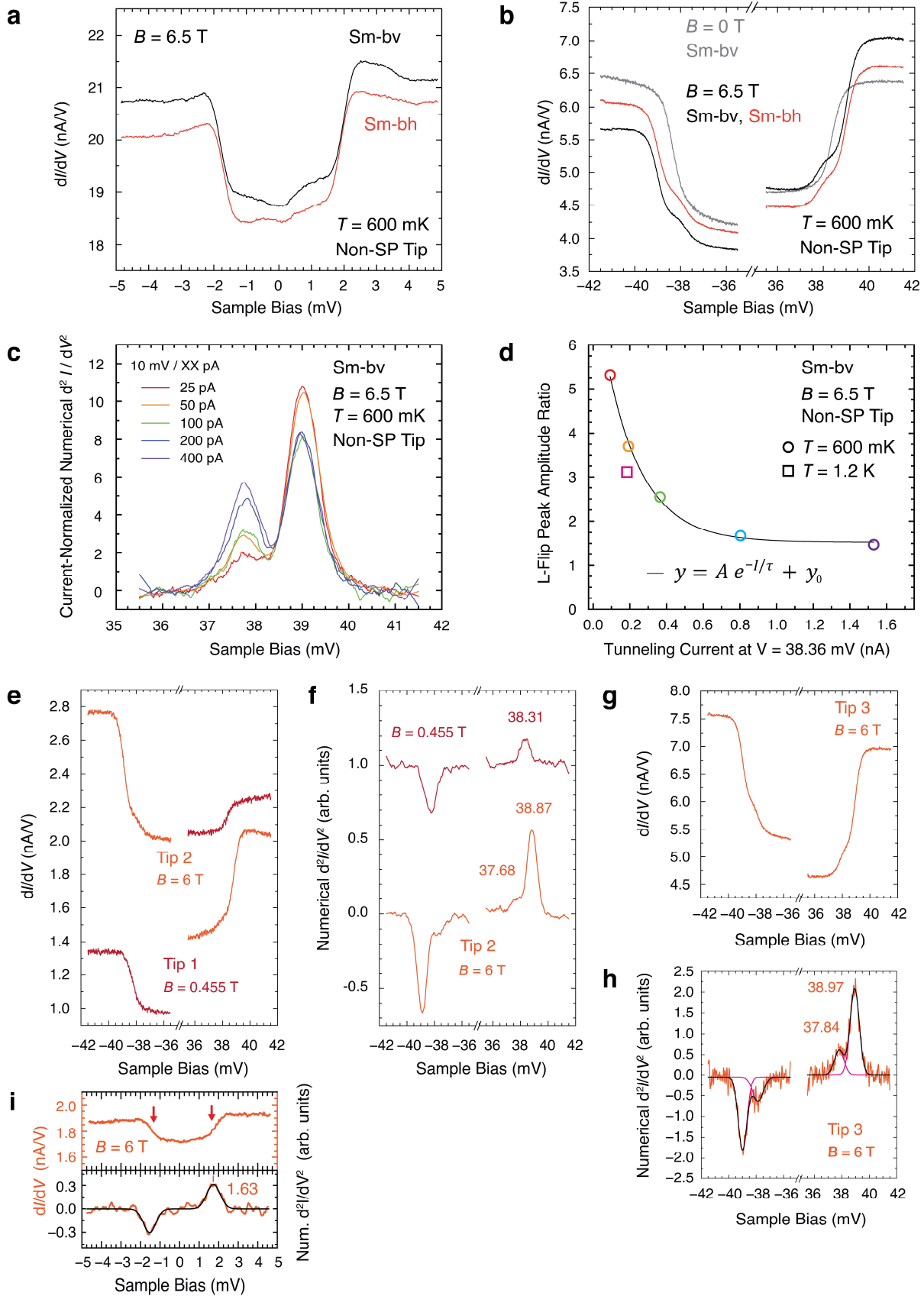

**Fig. S11. Tunneling spectra for Sm at bridge sites.**

**a**, Low-bias tunneling spectra (d$I$/d$V$) of bridge-site Sm atoms (Sm-bh and Sm-bv) show Zeeman splitting of the ground state at $B$ = 6.5 T. The numerical derivatives of these spectra (d$^2I$/d$V^2$) are shown in main text Fig. 4c. Anisotropy in the *g*-tensor, combined with the effect of the external magnetic field being more oriented along the O-O bridge direction for Sm-bh and more oriented along the Mg-Mg bridge direction for Sm-bv, results in the different energies for the two Sm-b orientations seen here and in Fig. 4c. A Gaussian fit to each peak in Fig. 4c, averaged over positive and negative bias, gives energies 1.786 ± 0.004 mV for Sm-bh and 1.893 ± 0.004 mV for Sm-bv. These energies correspond to g = 4.747 for Sm-bh and g = 5.031 for Sm-bv. Tip-height setpoint $I_{set}$ = 200 pA at $V_{set}$ = 10 mV, $V_{mod}$ = 0.2 mV rms. **b**, Tunneling spectra of Sm-b shows magnetic-field splitting of the ~38-mV peak. The numerical derivatives of these spectra (d$^2I$/d$V^2$) are shown in Fig. 4d. Splitting is symmetrical in energy but shows different peak heights. The effect of the bridge orientation (Sm-bv versus Sm-bh) is negligible. Tip-height setpoint $I_{set}$ = 50 pA at $V_{set}$ = 10 mV, $V_{mod}$ = 0.2 mV rms. **c**, Tip-height dependence of tunneling spectra of Sm-bv with tip height setpoint of 10 mV at current as labeled. Each spectrum is normalized by the tunneling current measured at 38.3 mV, which is the bias onset of the *L-S*-tilt excitation. **d**, Ratio of peak-heights of the split ~38 mV excitation from fits to 0.6 K data (open circles, spectra shown **c**) and at 1.2 K (open square, spectra not shown). **e**, Tunneling spectra (d$I$/d$V$) for Sm-b at $B$ = 0.455 T (dark red curves) and $B$ = 6 T (orange curves). The data was acquired with two different tip apexes as labeled, which both differed from the apex used for the spectra shown in **b** and Fig. 4d. Both sets of spectra used tip-height setpoint $I_{set}$ = 20 pA at $V_{set}$ = 10 mV, $V_{mod}$ = 0.2 mV rms. **f**, Corresponding numerical derivatives d$^2I$/d$V^2$ of spectra in **e**. The derivative used Savitzky-Golay smoothing in a 20-point window. **g**, Additional tunneling spectra for Sm-b acquired at $B$ = 6 T with a different tip apex ("Tip 3") and **h**, corresponding numerical derivative smoothed with a 20-point window. Data shown in **g** and **h** were acquired at setpoint $I_{set}$ = 50 pA at $V_{set}$ = 10 mV, $V_{mod}$ = 0.2 mV rms. **i**, Additional low-bias tunneling spectra (d$I$/d$V$, top) and corresponding numerical derivative (d$^2I$/d$V^2$, bottom) of Sm-bh showing the Zeeman splitting of the ground state doublet, acquired with a different tip apex than was used in **a** and Fig. 4c. Tip-height setpoint 200 pA at 10 mV, $V_{mod}$ = 0.2 mV rms.

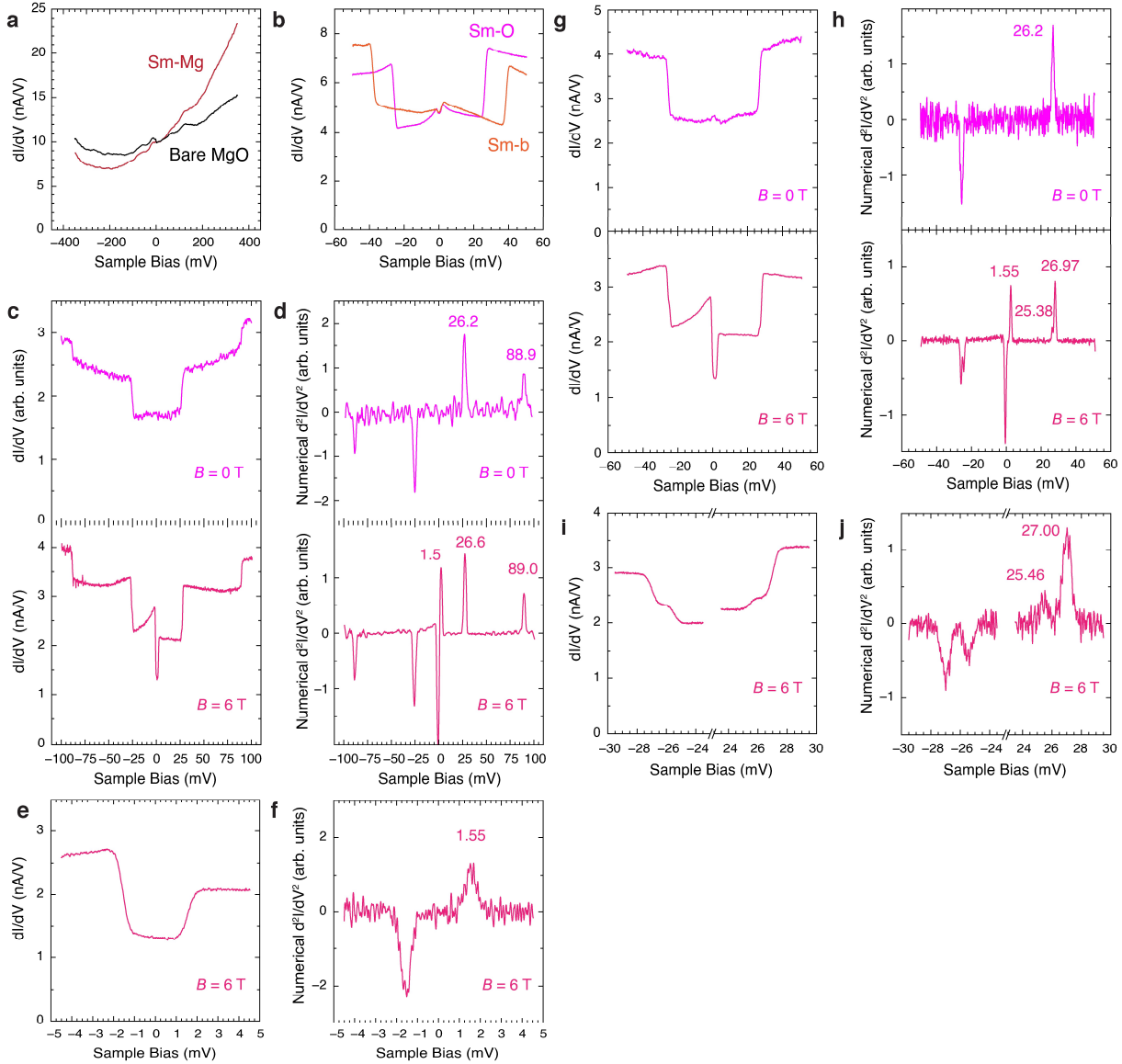

**Fig. S12. Tunneling spectra of Sm at Mg and oxygen sites.**

**a**, Tunneling spectra (d$I$/d$V$) acquired with the tip positioned over Sm-Mg (dark red) and the bare MgO substrate (black) using the same STM tip apex. Tip-height setpoint $I_{set}$ = 100 pA at $V_{set}$ = 10 mV. **b**, Tunneling spectra of Sm-b (orange) and Sm-O (pink). Setpoint $I_{set}$ = 10 pA at $V_{set}$ = 50 mV, $V_{mod}$ = 1 mV rms, B = 0.85 T. The same tip apex was used for both traces, but the apex used in **b** differs from that used in **a**. **c**, Tunneling spectra of Sm-O acquired at larger tip-surface distance and larger bias voltage range, showing the conductance step at ~89 mV. Acquired at $B$ = 0 T (top) with a non-spin-polarized tip and $B$ = 6 T (bottom) with a spin-polarized tip. Top spectrum: $I_{set}$ = 5 pA at $V_{set}$ = 20 mV, $V_{mod}$ = 1 mV rms. Bottom spectrum: $I_{set}$ = 5 pA at $V_{set}$ = 20 mV. **d**, Numerically computed d$^2I$/d$V^2$ spectra of corresponding traces in **c**. **e**, Low-voltage tunneling spectra measuring the Zeeman splitting of Sm-O near zero bias at $B$ = 6 T, acquired with the same spin-polarized tip as in **d** (bottom panel). $I_{set}$ = 20 pA at $V_{set}$ = 10 mV, $V_{mod}$ = 0.2 mV rms at $T$ = 0.6 K. **f**, Numerically computed d$^2I$/d$V^2$ spectrum of **e**. **g**, Zoom-in bias range at $B$ = 0 T (top) with a non-spin polarized tip and $B = 6$ T (bottom) with a spin-polarized tip. (top) $I_{set} = 25$ pA at $V_{set} = 10$ mV, $V_{mod} = 1$ mV rms at $T = 1.2$ K, (bottom) $I_{set} = 20$ pA at $V_{set} = 10$ mV, $V_{mod} = 0.3$ mV rms at $T = 1.2$ K **h,** Numerically computed $d^2I/dV^2$ spectra of corresponding traces in **g**. **i,** Tunneling spectra showing the splitting of the ~26 mV conductance step at $B = 6$ T. $I_{set} = 20$ pA at $V_{set} = 10$ mV, $V_{mod} = 0.2$ mV rms at $T = 0.6$ K. **j,** Numerically computed $d^2I/dV^2$ spectra of corresponding traces in **i**.

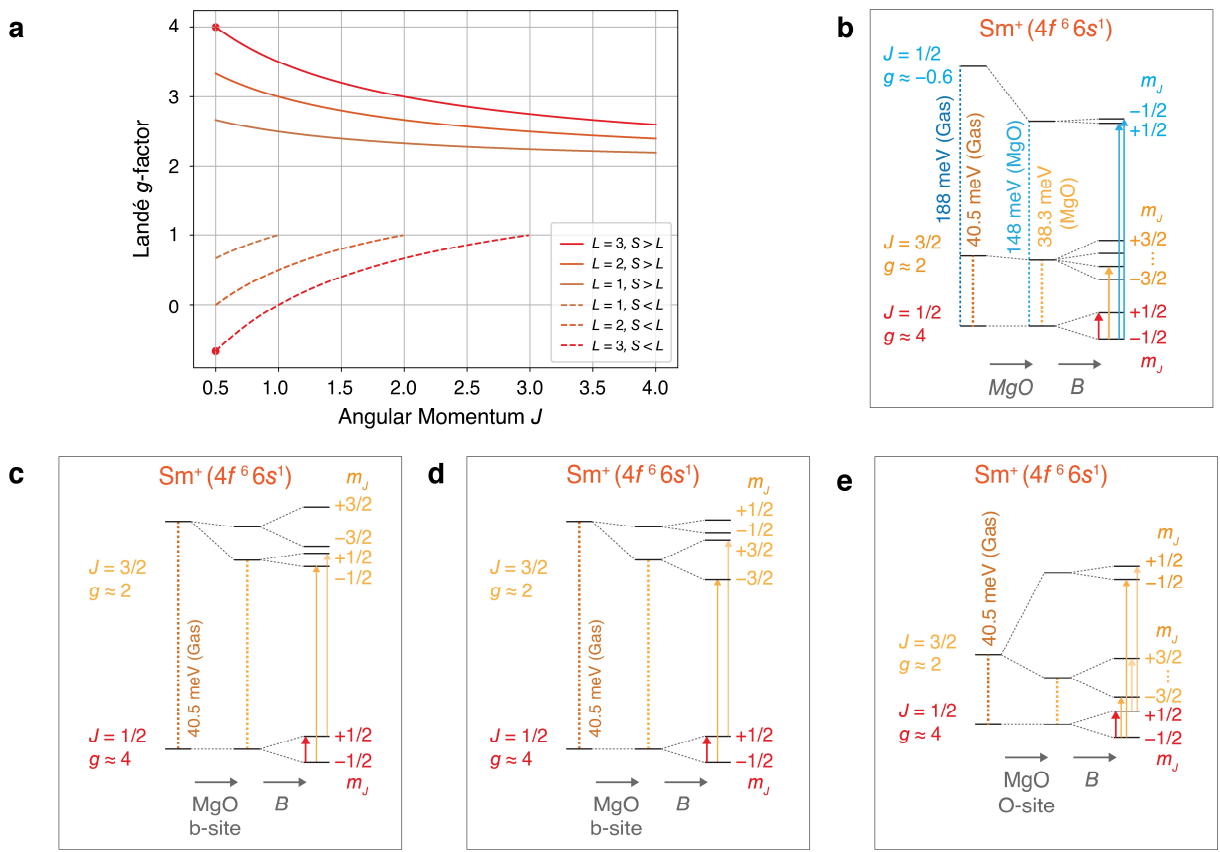

**Fig. S13. Proposed excitation schemes and calculated Landé *g*-factors for Sm$^+$**

**a**, Calculated Landé *g*-factors for various *L* as a function of *J*, including the case of *L* = 3 found in gas-phase Sm$^+$ (red). Red points correspond to the *L* = 3, *J* = 1/2 cases found in free Sm$^+$ ground and excited multiplets. Hund's third rule is assumed, so $J = |S - L|$. Lines show *J* varied continuously for illustration. Large *g*-factors occur when *J* is non-zero but minimized. (See Supplementary Note 7) **b**, Energy level scheme for gas-phase Sm$^+$, and for Sm at a bridge or oxygen site on MgO (Sm-b or Sm-O), with arrows showing IETS transitions, reproduced from main text Fig. 4e. Crystal field splitting is assumed to be near zero. **c,** Alternative excitation scheme in which the *J* = 3/2 excited states of Sm is split into two $m_J$ doublets by the crystal field when adsorbed on the b-site, with the $m_J$ = ±1/2 doublet lower in energy than the $m_J$ = 3/2 doublet. **d**, Alternative level scheme similar to **c** but with the $m_J$ = ±3/2 doublet lower in energy. **e**, Alternative excitation scheme for Sm adsorbed on the O-site in which the crystal field splitting is larger. (See Supplementary Note 5 for discussion of **b** to **d** and Supplementary Note 6 for **e**).

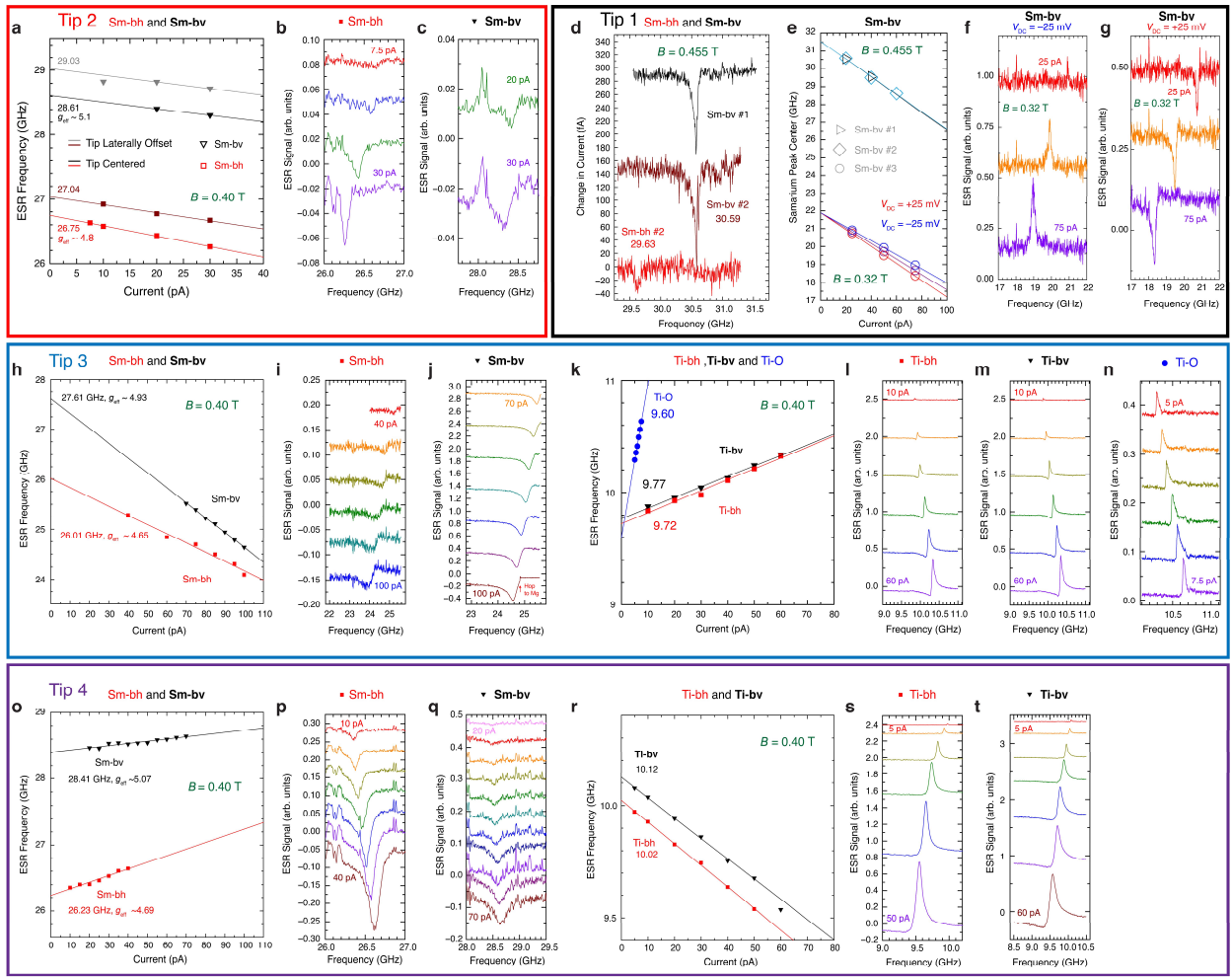

**Fig. S14. Anisotropic *g*-factor determination from ESR spectra of Sm-b.**

**a**, Tip-sample distance dependent ESR peak positions for Sm-bh (bridge-site, horizontal O-O orientation, squares) and Sm-bv (bridge-site, vertical orientation, triangles) acquired with the tip positioned laterally above the center of the atom (grey and brown points) or laterally offset ~0.15 nm (black and red points). All spectra acquired with $V_{DC}$ = 25 mV, $B$ = 0.40 T, and tip-height varied by varying the setpoint current $I_{set}$ as shown, at $V_{set}$ = 25 mV. **b**, Raw data for Sm-bh and **c**, raw data for Sm-bv, which yielded the red and black data points in **a** for the tip centered over the atom. All data in **a** to **c** was acquired with the same tip apex "Tip 2" and the same individual isolated Sm atom, which was repositioned between the bh and bv sites via atom manipulation. **d**, ESR spectra comparing Sm-bh and Sm-bv with a different tip apex, "Tip 1". **e**, Fitted ESR peak positions as a function of setpoint current for several Sm-bv atoms at different magnetic fields $B$. The data shown for $B$ = 0.32 T was acquired for both positive and negative bias polarity at the same tip-surface distance (open feedback loop after setting the tip height), demonstrating a subtle bias-voltage dependence on ESR frequency, an effect that has been noted recently for transition metal adatoms[20]. **f**, Raw data for spectra shown at $B$ = 0.32 T for negative bias and **g**, positive bias. The spectra in **f** and **g** have been normalized by the setpoint current. **h**, Additional ESR spectra comparing Sm-bh and Sm-bv with tip apex "Tip 3", along with raw data for Sm-bh **i**, and Sm-bv **j**, similar with **a** to **c**. All Sm spectra were acquired with bias $V_{DC}$ = +25 mV with $V_{set}$ = +25 mV and setpoint current as shown. **k**, Fitted ESR peak positions using the same tip apex "Tip 3" to acquire data on Ti-bh, Ti-bv and Ti-O atoms. **l–n**, Raw data for Ti-bh, Ti-bv and Ti-O, respectively, at different $I_{set}$ as labeled. Spectra vertically offset for clarity. All Ti spectra were acquired with bias $V_{DC}$ = +50 mV, $V_{set}$ = +50 mV and setpoint current as shown. **o–t,** Additional ESR spectra of Sm-bh, Sm-bv, Ti-bh, and Ti-bv acquired with a fourth tip apex, "Tip 4." All data for Sm and Ti were acquired under similar conditions as for "Tip 3" in **h–n**. All Sm spectra: $V_{RF}$ = 10 mV; all Ti spectra: $V_{RF}$ = 20 mV. Combining the extrapolated zero-current intercepts for tips 2–4 shown here gives an average effective *g*-factor of 5.04 ± 0.08 for Sm-bv and 4.71 ± 0.05 for Sm-bh.

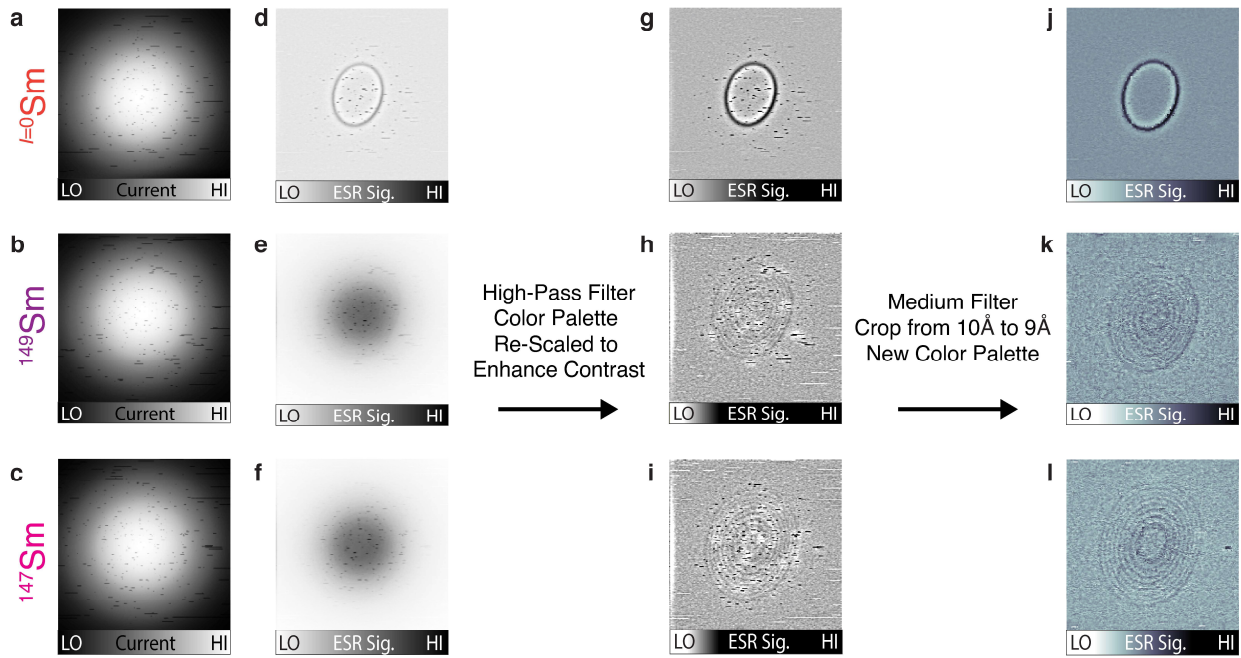

**Fig. S15. Post-processing of Sm magnetic resonance image (MRI) data.**
**a**–**c**, Tunneling current at constant tip height for $^{I=0}$Sm, $^{149}$Sm and $^{147}$Sm (respectively). The dark dots and lines ("popcorn" noise) resulted from a magnetically bi-stable STM tip and were not an intrinsic property of the Sm atoms. **d**–**f**, Raw magnetic resonance images (MRI) for each Sm isotope acquired simultaneously with the current maps shown in **a**–**c**. The ESR signal for $^{I=0}$Sm results in a single, clearly visible dark resonant slice that appears as an oval shaped ring. The smaller ESR signal for each resonant slice of the $I = 7/2$ nuclear isotopes (**e** and **f**) results in a series of concentric dark resonant slices which are not immediately obvious against the rectification background (diffuse dark region centered on the atom) or the popcorn noise. **g**–**i**, High-pass filtered data from **d**–**f** rescaled non-linearly to emphasize the contrast arising from the ESR signal. Here the high-pass filtering had the effect of removing the rectification background which varied slowly with lateral tip position while preserving the ESR signals which varied rapidly with tip position. **j**–**l** Cropped, median filtered images plotted with an alternative color palette that was similarly adjusted to emphasize contrast (non-linear). Here median filtering was used to remove the "popcorn" noise arising from switching of the tip magnetization during image acquisition. Each image was cropped from 1 nm to 0.9 nm to remove the imaging artifact at the start of each horizontal scan line (bright edge seen on the left of each image in **g**–**i**).